\documentclass[preprint]{aastex631}
\usepackage{CJK}
\usepackage{amsfonts}
\usepackage{amsmath}
\usepackage{mathrsfs}
\usepackage{epsfig}
\usepackage{subfigure}

\newcommand\diff{\,\mathrm{d}}

\def\m{\,\mathrm{m}}

\def\s{\,\mathrm{s}}
\def\p{\,\mathrm{p}}

\def\C{\,\mathrm{C}}
\def\N{\,\mathrm{N}}
\def\O{\,\mathrm{O}}

\def\B{\,\mathrm{B}}
\def\A{\,\mathrm{A}}

\def\c{\,\mathrm{c}}

\def\g{\,\mathrm{g}}
\newcommand\TESS{{\it TESS }}
\def\BJD{\,\mathrm{BJD}}
\def\TDB{\,\mathrm{TDB}}
\def\P{\,\mathrm{P}}
\def\Teff{\,T_{\mathrm{eff}}}
\def\dPdt{\,(1/P_0)(\diff P_0/\diff t)}
\def\yr{\,\mathrm{yr}}
\def\cd{\,\mathrm{c}\ \mathrm{d}^{-1}}
\def\logg{\,\log{g}}
\def\Qobs{\,Q_\mathrm{obs}}
\def\Qmod{\,Q_\mathrm{mod}}
\def\raa{Res.\ Astron.\ Astrophys.\ }

\shorttitle{Evolutionary Asteroseismology of AE UMa}
\shortauthors{H.-F. Xue, J.-S. Niu \& J.-N. Fu}

\begin{document}
\begin{CJK*}{UTF8}{gbsn}

\title{Precise Evolutionary Asteroseismology of High-Amplitude $\delta$ Scuti Star AE Ursae Majoris}

\correspondingauthor{Jia-Shu Niu}
\email{jsniu@sxu.edu.cn}

\author[0000-0001-6027-4562]{Hui-Fang Xue (薛会芳)}
\affil{Department of Physics, Taiyuan Normal University, Jinzhong 030619, China}
\affil{Institute of Computational and Applied Physics, Taiyuan Normal University, Jinzhong 030619, China}

\author[0000-0001-5232-9500]{Jia-Shu Niu (牛家树)}
\affil{Institute of Theoretical Physics, Shanxi University, Taiyuan 030006, China}
\affil{State Key Laboratory of Quantum Optics and Quantum Optics Devices, Shanxi University, Taiyuan 030006, China}

\author[0000-0001-8241-1740]{Jian-Ning Fu (付建宁)}
\affil{Department of Astronomy, Beijing Normal University, Beijing 100875, China}

\begin{abstract}
Stellar structure and evolution theory is one of the basis in modern astronomy.
   Stellar inner structures and their evolutionary states can be precisely tested by asteroseismology, since the inner information is brought to the stellar surface by the global oscillating waves and becomes observable.
   For stellar evolutionary speed (i.e. how long time scale does a star stay at a special evolution phase?), because of the insurmountable gap between the time scales of the evolutionary history of human civilization and a star, it can only be roughly tested by ensemble of stars in different evolutionary stages in most cases, and all the snapshots of these stars make up our global view of stellar evolution. 
  The effect of stellar evolution on the structure and the corresponding global size of a pulsating star will lead to tiny period variations of its pulsation modes, which are the most valuable indicators of its evolutionary state and can be used to test the stellar evolution theory by a single star rather than ensemble of stars.
  Here, we report a High-Amplitude $\delta$ Scuti star AE Ursae Majoris, who locates in the post main-sequence (MS) evolutionary stage and its observed linear period variation rate can be practically ascribed to its evolutionary effect.
  The result tests the stellar evolution theory from the pre-MS to post-MS with an unprecedented precision by a single star, and the framework can be extended to other type of pulsating stars to perform precise evolutionary asteroseismology, which aims to test the current stellar evolution theory in different evolutionary stages, discover the discrepancies between the theory and observations, and ultimately build a complete and precise stellar evolution theory to backtrack the history of each of these stars.
\end{abstract}

\section{Introduction}           

$\delta$ Scuti stars are a kind of short-period pulsating variable stars whose periods range from 15 minutes to 8 hours and the spectral classes A-F. They locate on the main sequence (MS) or post-MS evolutionary stage at the bottom of the classical Cepheid instability strip in Hertzsprung-Russell diagram (H-R diagram) and are self-excited by the kappa mechanism due to the partial ionization of helium in the out layers \citep{Breger2000,Handler2009,Uytterhoeven2011,Holdsworth2014}. High-amplitude $\delta$ Scuti stars (hereafter HADS) are a kind of $\delta$ Scuti stars, who have relatively larger amplitudes ($\Delta V \geq 0_{\cdot}^{m}3$) and slower rotations in most cases. Most of HADS show single or double radial pulsation modes, and some of them have three radial pulsation modes or even some non-radial pulsation modes \citep{Niu2013,Niu2017,Xue2018,Wils2008,Xue2020,Daszynska2022}.

Limited by the precision of ground-based photometric observations, only the linear period variation rates of the pulsation modes with the highest amplitudes (usually the fundamental modes) of HADS can be obtained by the accumulation of the data lasting for decades \citep{Yang2012,Niu2017,Xue2018}.
The high-precision continues photometric observations from spacecraft like {\it Kepler} (lasting for four years) can also lead us to get the linear period variation rates of HADS, not only for the highest amplitude pulsation modes, but also for other pulsation modes with sufficient signal-to-noise ratio \citep{Bowman2016,Niu2021,Bowman2021}.
These linear period variation rates ($(1/P)(\diff P/\diff t)$) generally have values $\pm (10^{-9} - 10^{-6}) \yr^{-1}$.
On the aspect of stellar evolution theory, HADS are always considered to locate in the main-sequence (MS) or post-MS evolutionary stages.\footnote{HADS are also have the possibility in the stage of pre-MS, which should have larger and negative values of the linear period variation rates.} In these stages, the linear period variation rates\footnote{Here, we refer to the radial pulsation modes, especially the fundamental pulsation mode in most cases.} caused by the stellar evolution cannot exceed $20 \cdot 10^{-8}\ yr^{-1}$ \citep{Breger1998}, which are systematically smaller than the observed values and indicate that there exist additional mechanisms (such as mass transfer, light traveling time effect in multiple system, and nonlinear mode interactions) contributing to the linear period variation rates \citep{Breger1998,Bowman2016,Bowman2021,Xue2020}. However, some recent researches \citep{Niu2017,Xue2018} show that the observed linear period variation rates of some HADS can be roughly included in their asteroseismological models self-consistently.

Whether the observed linear period variation rates of HADS (at least in some individual cases) can be fully predicted by the stellar evolution theory is directly related to two important issues:
(i) the physical essence of HADS: the low-amplitude $\delta$ Scuti stars (LADS) always have similar pulsation periods and seem to have almost the same basic physical properties (such as the effective temperature, the luminosity and the surface gravitational acceleration) as HADS, but they have obvious different light curve features. What's the origin of the difference? If it comes from the different stellar evolution stages \citep{Petersen1996}, one should precisely determine the evolutionary stages of them in the framework of current stellar evolution theory, which requires at least the consistency of the observed linear period variation rates and the stellar evolution theory predictions.
(ii) the stellar evolution theory: because of the insurmountable gap between the time scales of the evolutionary history of human civilization and stars, the stellar evolution theory is generally constructed, tested, and updated by ensemble of stars in different evolutionary stages base on a statistical sense.
The comparison of the linear period variation rates between observations and theoretical predictions of pulsating stars in a rapidly evolving stage (such as HADS) provides us an opportunity to test the current stellar evolution theory by single stars precisely.
The deviations between them are the indicators of the self-consistency of the stellar structure and evolution theory to some extend, which would be the important hints to update the current theory in future.

AE Ursae Majoris (hereafter AE UMa, $\alpha_{2000} = {09^h}{36^m}{53^s}$, $\delta_{2000} = 44^\circ04^\prime01^{\prime\prime}$, V=11.27 mag) is a Population I, post-MS HADS pulsating in radial modes \citep{Hintz1997, Pocs2001, Zhou2001, Niu2017}. Based on the detected frequencies of the fundamental and first overtone pulsation modes ($f_0=11.6256\ \cd$, $f_1=15.0312\ \cd$), together with the linear period variation rate of the fundamental pulsation mode ($\dPdt = (5.4 \pm 1.9) \times 10^{-9}\ \yr^{-1}$), \citet{Niu2017} has made a pioneering attempt to do asteroseismology for AE UMa. It shows that AE UMa could be a normal star evolving into the post-MS stage, although the pulsating frequencies obtained from the ground-based telescopes have relatively large uncertainties and the final results have obvious deviations comparing with the observed spectroscopic parameters.

These years, AE UMa has been monitored not only by the Transiting Exoplanet Survey Satellite (\TESS \citep{Ricker2015}), but also by some other ground-based observations whose data have been collected by AAVSO\footnote{American Association of Variable Star Observers, https://www.aavso.org}. The continues photometric data from \TESS provide us accurate pulsating frequencies, and all the extended data accumulate the number of the times of maximum light (TML), which are necessary for determining a reliable value of the observed linear period variation rate of the fundamental pulsation mode.

\section{Methods}

\subsection{Frequency Analysis}

\begin{figure*}[htp]
   \centering
   \includegraphics[width=0.95\textwidth, angle=0]{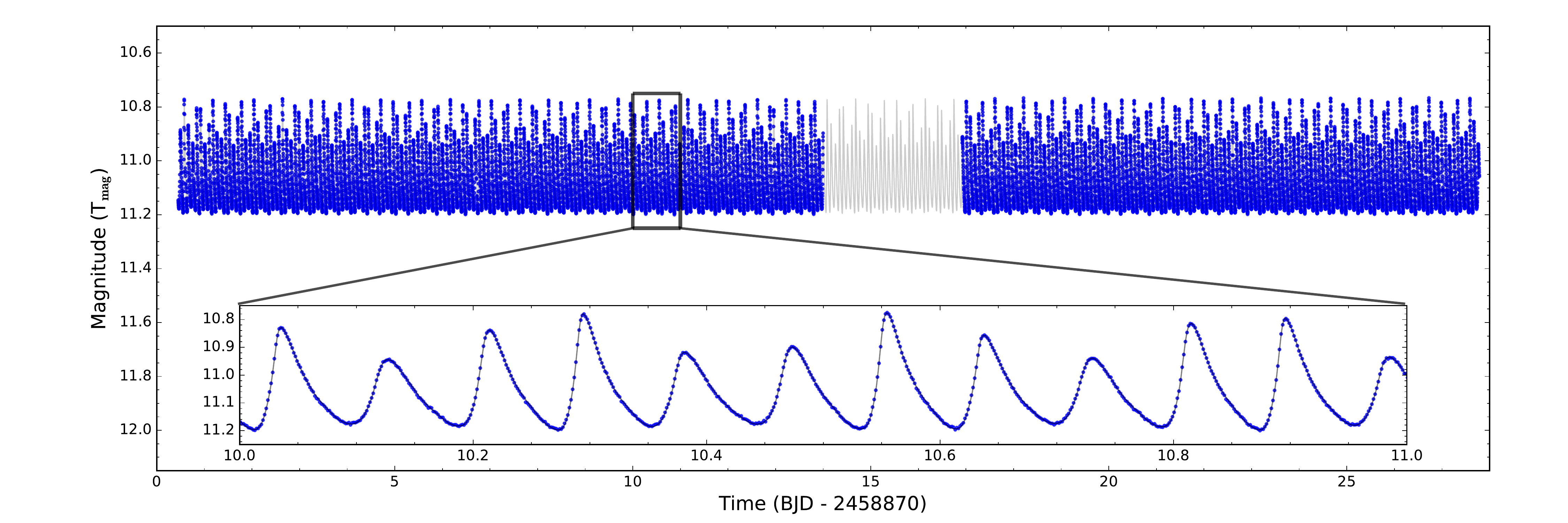}
   \caption{Light curves of AE UMa observed by TESS (blue points) and fitted by 65 extracted frequencies (gray line). The small panel is a zoom.}
   \label{fig:lc}
\end{figure*}

AE UMa (TIC 357132618) has been observed by \TESS spacecraft during Sector 21 from 2020 January 21 to 2020 February 18 (BJD 2458870.45 - 2458897.78), in which there exists a gap of about 4 days (see in Figure \ref{fig:lc}). We downloaded the 2-min cadence flux measurements from MAST Portal\footnote{https://mast.stsci.edu/portal/Mashup/Clients/Mast/Portal.html} which were processed by the \TESS Science Processing Operations Center (SPOC; \citet{Jenkins2016}). After converting the normalized fluxes to magnitudes by utilizing a \TESS magnitude of +11.089 \citep{Stassun2019}, we totally got 17336 data points. The light curves are shown in Figure \ref{fig:lc}.

\begin{figure*}[htp]
   \centering
   \includegraphics[width=0.95\textwidth]{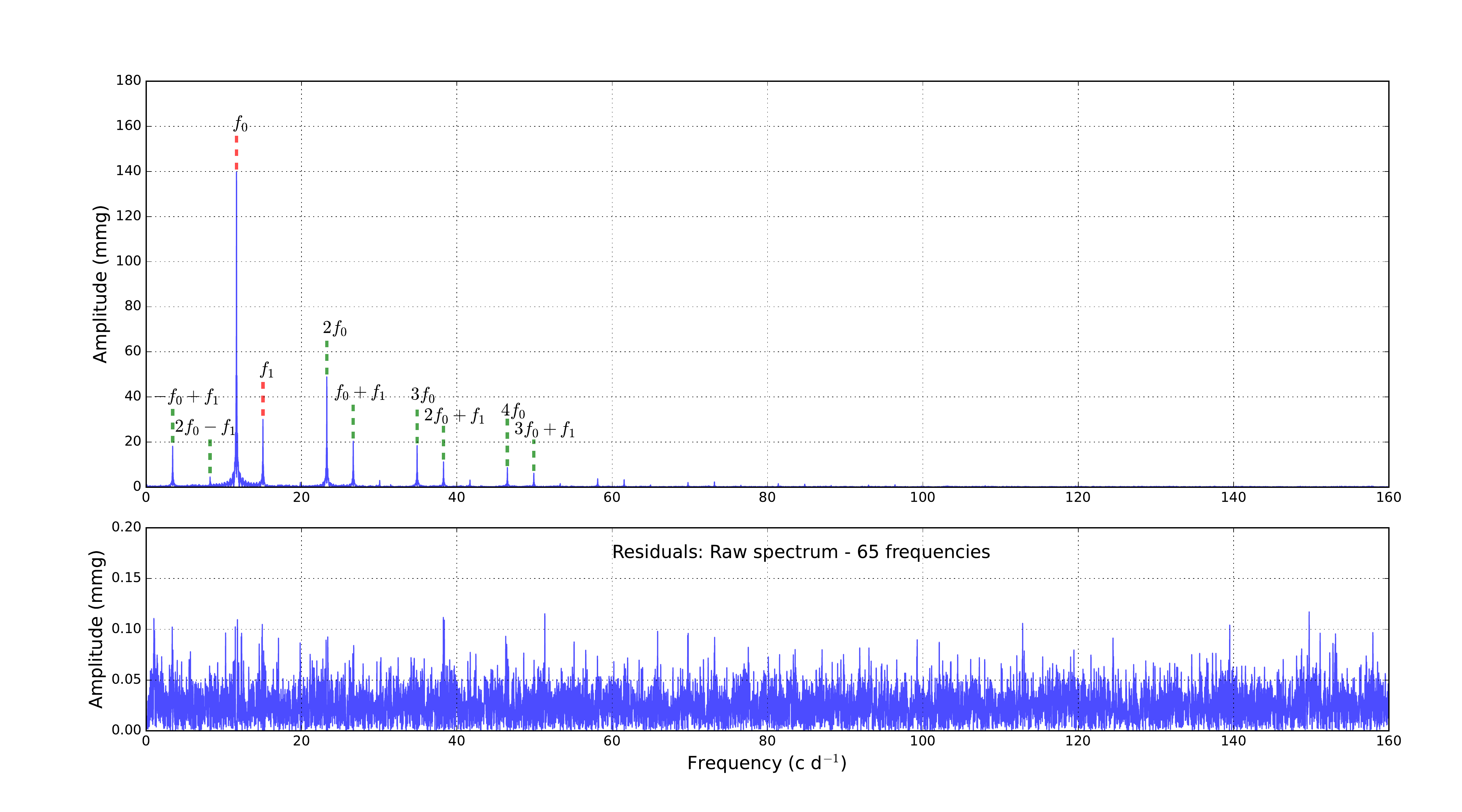}
   \caption{Frequency spectrum before (top panel) and after (bottom panel) the pre-whitening process. All the frequency peaks larger than 5 mmag are marked in the top panel.}
   \label{fig:fre_spe}
\end{figure*}

In this work, the pre-whitening process was performed to the light curves observed by \TESS based on the framework introduced in \citet{Zong2016,Niu2021}, in which the uncertainties of amplitudes ($\sigma_a$) were defined as the median value of the amplitudes within a Lomb-Scargle spectral window of 2 $\cd$ equally divided by the frequency peak, while the uncertainties of frequencies ($\sigma_f$) were estimated following the formalism proposed by \citet{Montgomery1999,Aerts2021}.
The statistical significant criterion of the signal-to-noise in the frequency spectrum was set to be 4.0 \citep{Breger1993}, and finally, 65 frequencies were extracted, which were composed by two independent frequencies ($f_{0} = 11.6257 \pm 0.0002\ \cd$ and $f_{1} = 15.0314 \pm 0.0002\ \cd$) and their harmonics and combinations. Figure \ref{fig:fre_spe} shows the origin frequency spectrum and the residues after extracting the 65 significant frequency solutions. All the significant frequency solutions are listed in Appendix, Table \ref{tab:fre_solu}.

The period ratio ($P_{1}/P_{0} = f_{0}/f_{1}$) has a value of 0.7734, which agrees with the theoretical calculated value ($\sim0.77$, \citet{Petersen1996,Poretti2005}) and proves that these periods belong to the radial fundamental and first overtone pulsation modes.

\subsection{Linear Period Variation Rate}

The classical $O-C$ method was used to obtain the linear period variation rate of the fundamental pulsation mode.
The TML were collected from the historical literatures, AAVSO, and \TESS.

We collected 443 TML from \citet{Niu2017}, and 23 more from the subsequent literatures. 
Based on the photometric data from AAVSO (from 1998 February to 2020 April) and  \TESS (from 2020 January 21 to 2020 February 18), we obtained the TML by a fourth polynomial fitting around the maxima in the light curves, whose uncertainties were estimated by Monte Carlo simulations. In this process, we determined 244 and 287 TML from AAVSO and \TESS, respectively. All these TML except that from \TESS were transformed into Barycentric Julian Dates based on the Barycentric Dynamical Time standard ($\rm{BJD_{TDB}}$) by using an online applets\footnote{http://astroutils.astronomy.ohio-state.edu/time/} \citep{Eastman2010}. Finally, we had collected 997 TML lasting for about 46 years to perform the $O-C$ analysis, and all of them are listed in Appendix, Table \ref{tab:O-C}. Those TML without giving uncertainties in the sources were uniformly estimated to give uncertainties based on their least significant digits from the sources.

A quadratic polynomials was used to reproduce the TML:
\begin{equation}
\BJD_{\TDB}=\BJD_{0} + P_{0} \cdot E + \frac{1}{2} \beta \cdot E^{2} ,
\label{eq:quadratic}
\end{equation}
where $\BJD_{\TDB}$ is the TML, $\BJD_0$ is the reference epoch, $P_0$ is the period of the fundamental pulsation mode, $\beta$ is the linear change of pulsation period, and $E$ is the epoch.

\begin{figure*}
   \centering
   \includegraphics[width=0.95\textwidth]{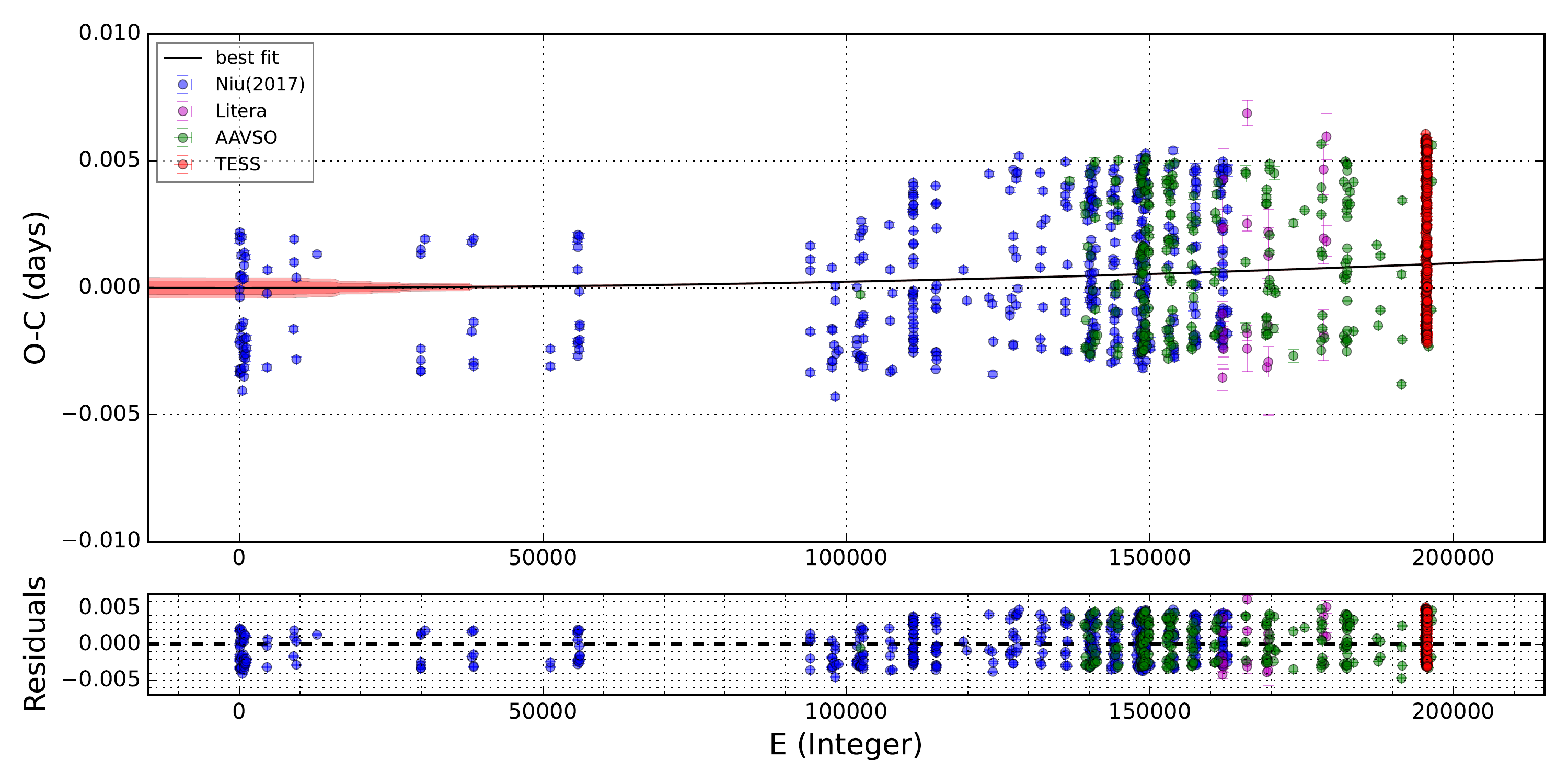}
   \caption{$O-C$ values and the corresponding residuals. In the upper panel, the black line represents the best-fit result of a quadratic fitting, the circles with different colors represent the different data sources. In the lower panel, the residuals of the best-fit result are plotted. The data collected from \citet{Niu2017} are shown in blue points; the data from historical literatures are shown in purple points; the data from AAVSO are shown in green points; the data from \TESS are shown in red points. The $2\sigma$ (deep red) and $3\sigma$ (light red) bounds of the fitting results are also shown in the figures.}
   \label{fig:O-C}
\end{figure*}

The Markov Chain Monte Carlo (MCMC) algorithm was used to determine the posterior probability distribution of the parameters in Eq. (\ref{eq:quadratic}).\footnote{The {\sc python} module {\tt emcee} \citep{emcee} was employed to do the MCMC sampling. Some of the examples can be referred to \citet{Niu201801,Niu201802,Niu2019} and references therein.} After the Markov Chains had been stable, the samples of the parameters were taken as their posterior probability distribution function (PDF) to obtain the mean values and the standard deviation of the parameters ($\BJD_{0} = 2442062.58391 (\pm 0.00001)$, $P_{0} = 0.0860170616 (\pm 0.0000000002)$, $\beta = 6.0 (\pm 0.1) \times 10^{-14}$). The best-fit result of the $O-C$ values and the corresponding residuals are shown in Figure \ref{fig:O-C}. Based on the fitting results of $P_0$ and $\beta$, the linear period variation rate can be obtained as $\dPdt=(2.96 \pm 0.05) \times 10^{-9}\ \yr^{-1}$.
Here, one should note that the uncertainty of the value of $\dPdt$ decreases by a factor of about 40 compared with our previous work ($(5.4 \pm 1.9) \times 10^{-9} \yr^{-1}$ in \citet{Niu2017}), which comes from both the additional TML and the different uncertainty estimation methods in these works.

\subsection{Theoretical Models}

The stellar evolution code Modules for Experiments in Stellar Astrophysics (MESA; \citet{Paxton2011,Paxton2013,Paxton2015,Paxton2018,Paxton2019}) was used to build the structure and evolution models. The stellar oscillation code GYRE \citep{Townsend2013,Townsend2018,Goldstein2020} was used to compute the corresponding frequencies of the pulsation modes for a specific structure model.

According to the previous work of \citet{Niu2017}, we fix the initial helium mass fraction, metal mass fraction, and the mixing-length parameter as $Y=0.25$, $Z=0.008521$, and $\alpha_{MLT}=1.89$, respectively. In addition, the mass ranges from 1.50 $M_\odot$ to 2.50 $M_\odot$ with a step of 0.01 $M_\odot$.
 Considering AE UMa as an HADS with very slow rotation \citep{Breger2000}, the effect of rotation on the structure and evolution models were disregarded. Each of the evolutionary tracks were calculated from pre-MS to post-MS evolutionary stages. At every step in the evolutionary tracks, the frequencies of multiple pulsation modes were calculated.

The exponential overshooting scheme \citep{Herwig2000} was tested with different $f_{\mathrm{ov}}$ values (obtained from \citet{Magic2010}). However, the best-fit with the the minimum value of $\chi^{2}$ (obtained by $f_{0}$ and $f_{1}$) results not only cannot give out effective temperature and luminosity values within uncertainties, but also showed large deviations to the observed linear period variation rate (exceeding 160\%). 
As a result, the effect of overshooting were not included in our evolutionary models.

\section{Results and Discussions}
\begin{figure*}[htp]
  \centering
    \includegraphics[width=0.6\textwidth]{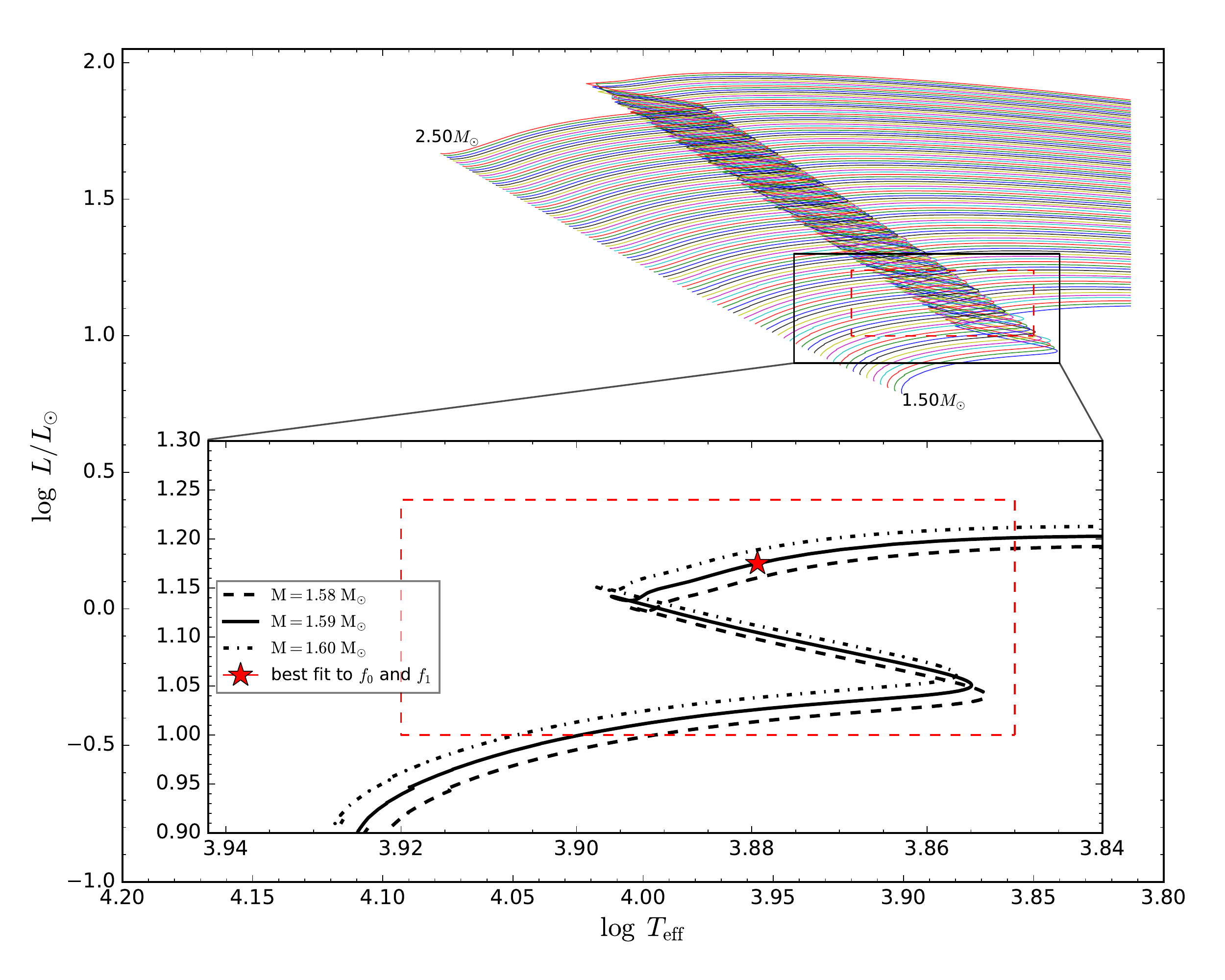}
    \includegraphics[width=0.6\textwidth]{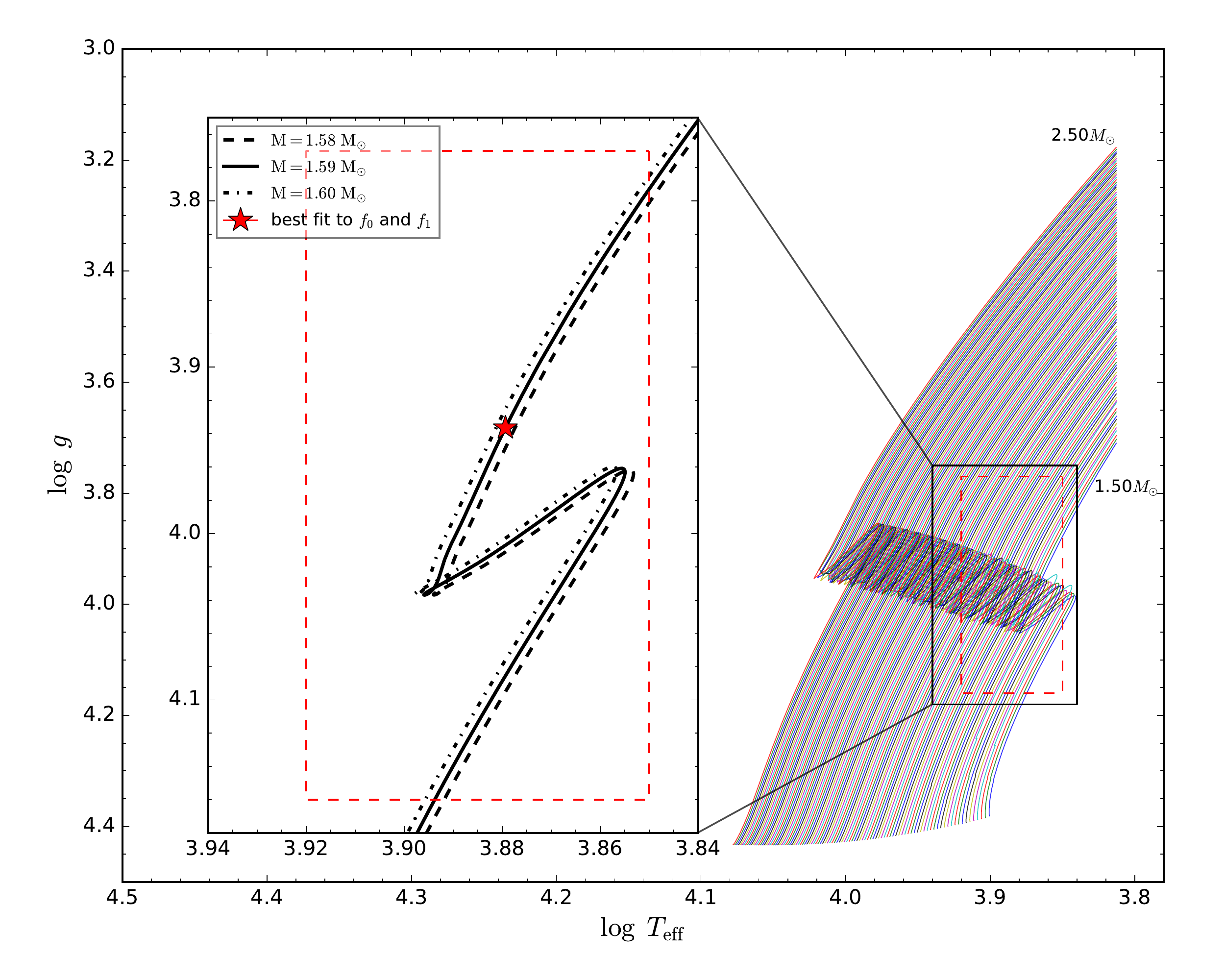}
    \caption{Hertzsprung-Russell diagram and $\log \Teff$ vs $\log g$ diagram with all the evolutionary tracks and that of the best-fit model. The colored evolutionary tracks show from the zero-age MS to post-MS evolutionary stage, with initial mass from $1.50\ M_{\odot}$ to $2.50\ M_{\odot}$. The regions surrounded by the black rectangular boxes are selected to zoom in the observed constraints and the best-fit model, which are represented by the red dashed rectangular boxes and the red stars respectively.}
  \label{fig:spectra_info}
\end{figure*}

\begin{figure*}[htp]
  \centering
    \includegraphics[width=0.6\textwidth]{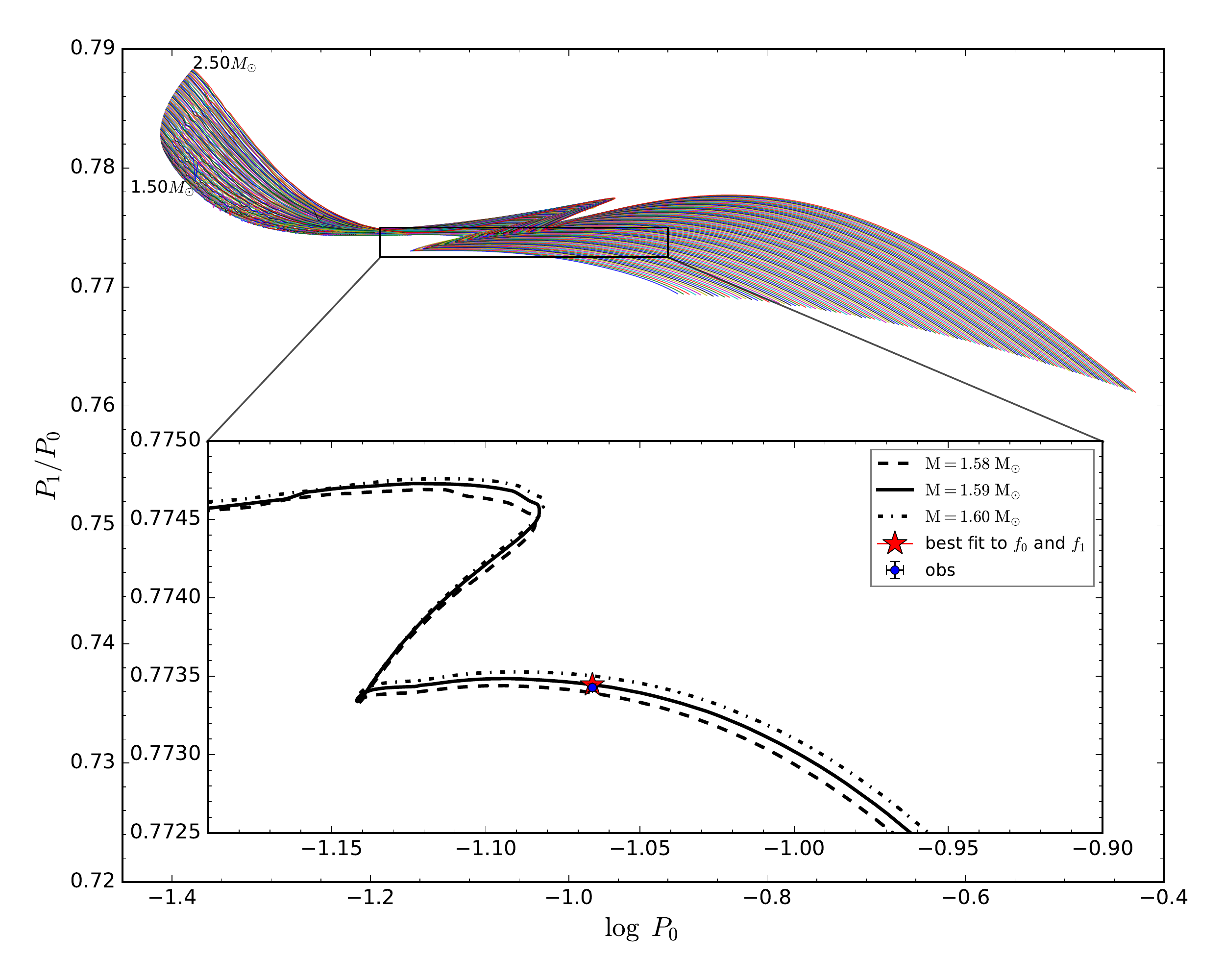}
    \includegraphics[width=0.6\textwidth]{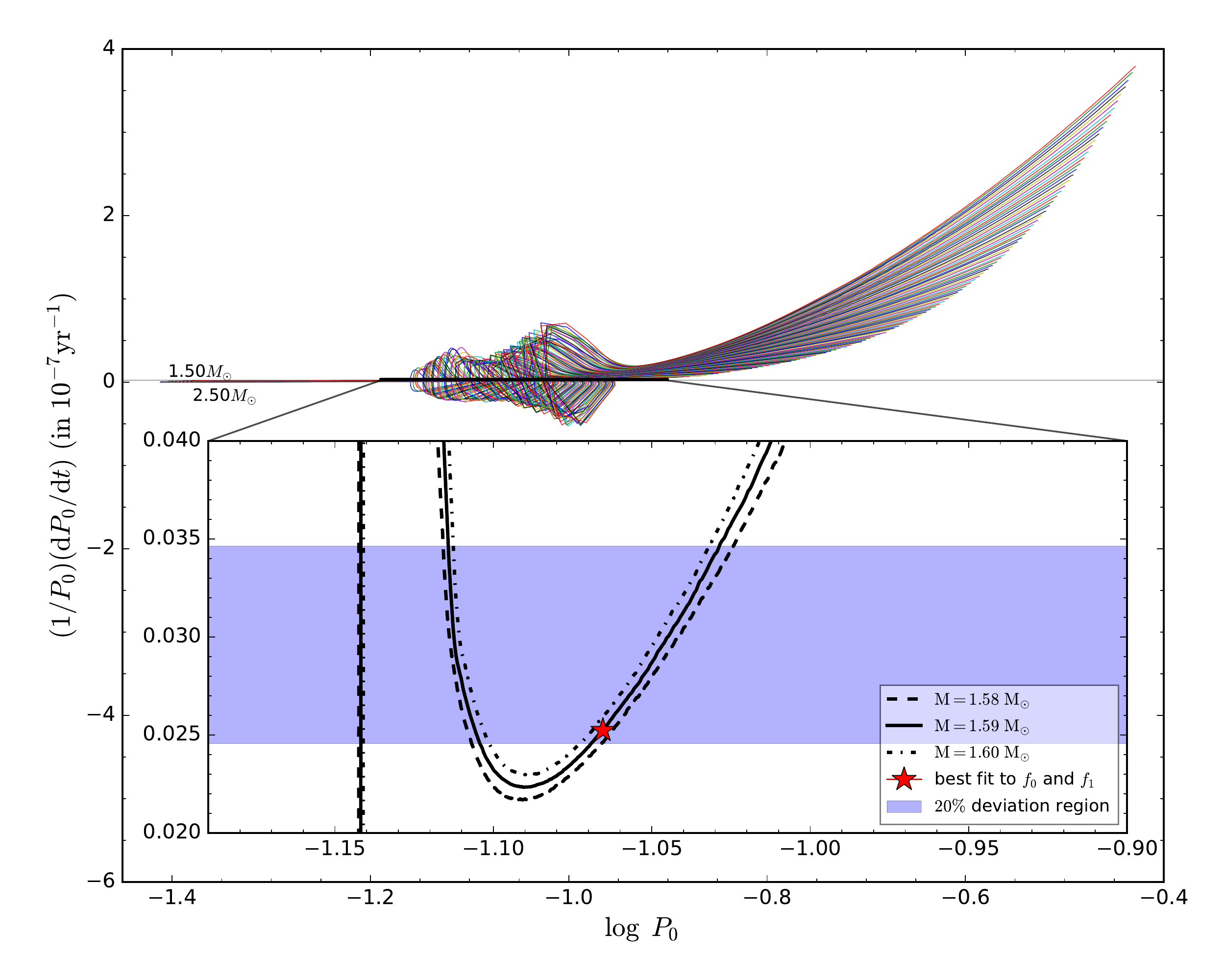}
    \caption{Petersen diagram and $\log P_{0}$ vs $\dPdt$ diagram with all the evolutionary tracks and that of the best-fit model. The colored evolutionary tracks show from the zero-age MS to post-MS evolutionary stage, with initial mass from $1.50\ M_{\odot}$ to $2.50\ M_{\odot}$. The regions surrounded by the black rectangular boxes are selected to zoom in the best-fit model, which are represented by the red stars. In the $\log P_{0}$ vs $\dPdt$ diagram, the blue region represents the 20\% deviation region of the observed $\dPdt$. }
  \label{fig:astero_info}
\end{figure*}

The best-fit model was determined by the observed fundamental and first overtone frequencies, which gave out the minimum value of $\chi^{2}$. Some of the important physical parameters of the best-fit model are listed in Table \ref{tab:params_all}.  
The Hertzsprung-Russell diagram and $\log \Teff$ vs $\log g$ diagram with all the evolutionary tracks and the best-fit model are shown in Figure \ref{fig:spectra_info}, which represent that the effective temperature, the luminosity and the surface gravitational acceleration are all well matched between the observations and the theoretical predictions.\footnote{The luminosity was calculated by the Gaia astrometric data and some other physical parameters were collected from the previous works. All these observed parameters are collected in Table \ref{tab:params_all}. See more information about the details in Appendix.}
The Petersen diagram and $\log P_{0}$ vs $\dPdt$ diagram are shown in Figure \ref{fig:astero_info}, which represent that not only the observed $P_1/P_0$ and $\log{P_0}$ match well with the calculated ones when the stellar mass equals to 1.59 $M_\odot$, but the linear period variation rate of the best-fit model also falls into the $20 \%$ deviation region of the observed values.

\begin{table*}[htp]
  \caption{Some of the important parameters from observation and the best-fit model. The deviation in the last column is defined as $\vert (\Qobs - \Qmod)/\Qmod \vert$, where $\Qobs$ is the observed quantity and $\Qmod$ is the corresponding quantity from the theoretical model calculation.}
  \centering
  \begin{tabular}{c|cc|c}
    \hline
    \hline
    Parameters & Observed& Best-fit& Deviation\\
    \hline
    $\Teff\ \mathrm{(K)} $    & $[7150, 8320]$ & $7575$ & --- \\
    $\log L/L_{\odot}$   & $[1.00, 1.24]$ & $1.18$  & ---\\
    $\log g$   & $[3.77, 4.16]$ & $3.94$ & ---\\
    \hline
    $f_{0}\ (\cd) $   & $11.6257 \pm 0.0002$ & $11.6259$ & $0.0017\%$\\
    $f_{1}\ (\cd) $   & $15.0314 \pm 0.0002$ & $15.0313$ & $0.0007\%$\\
    $\dPdt\ (\yr^{-1})$   & $(2.96 \pm 0.05) \times 10^{-9}$ & $2.52 \times 10^{-9}$ & $ 17\%$\\
    \hline
    $\mathrm{Mass}\ (M_{\odot})$ & --- & $1.59$  & --- \\
    $\mathrm{Age\ (Gyr)}$  & --- & $1.56$ & --- \\
    $\log R/R_{\odot}$ & --- & $0.35$ & --- \\
    \hline
  \end{tabular}
  \label{tab:params_all}
\end{table*}

These results prove that the observed linear period variation rate can be practically ascribed to the stellar evolution for AE UMa, which means that its evolutionary stage (post-MS) can be precisely determined in the framework of current stellar evolution theory. If we confirmed the evolutionary stages of more HADS and LADS in such way, we might have the answer to the question: why their light curves are so different from each other?\footnote{Here, we choose to avoid the HADS and LADS whose linear period variation rates are obvious larger than that caused by the stellar evolution, which could come from mass transfer, light traveling time effect in multiple system, and nonlinear mode interactions. These are related to other topics and not discussed in this work.}
On the other hand, the stellar evolution theory is precisely tested from the pre-MS to post-MS evolutionary stages within a deviation less than 20\% by a single star rather than ensemble of stars. Although the 17\% deviation could come from both the observation and theoretical model, it is a potential hint for updating the stellar evolution theory in future.

Although the concept of high-precision asteroseismology has been proposed and pioneeringly performed on some multi-modes pulsating stars (such as solar-like oscillation stars \citep{Wu2016,Wu2017}, slowly pulsating B stars \citep{Wu2019}, and red-giant branch stars \citep{Zhang2020}) to precisely probe the stellar inner structure and test input physics, the involvement of period variation makes it possible for us to determine the evolutionary status of some pulsating stars (such as HADS \citep{Niu2017,Xue2018} and blue large-amplitude pulsators \citep{Wu2018,Pigulski2022}).
The observed period variation of the pulsating stars can be contributed by different mechanisms, such as stellar evolution \citep{Niu2017,Xue2018}, mass transfer \citep{Xue2020}, light traveling time effect in a multiple system \citep{Xue2020}, and nonlinear mode interactions \citep{Breger2014,Zong2016,Bowman2016}, which have different time scales. Generally speaking, the stellar evolution has the longest time scale (like the case of white dwarfs: \citet{Kepler2021}). If we selected some pulsating stars whose observed linear period variation rates falls into an interval derived from the theoretical predictions, we could have the opportunity to test and update the stellar evolution theory precisely by these single stars via the framework in this work.
Here, we would like to propose the concept of Precise Evolutionary Asteroseismology (PEA).
In PEA, we pay special attention to the above pulsating stars in different evolutionary stages. According to the comparison of the observed linear period variation rates and the theoretical predicted ones (according to asteroseismology), we can test the self-consistency of current stellar structure and evolution theory precisely in different evolutionary stages. The discrepancies can be used to guide further improvements or updates of the theory.
At last, PEA will bring us a complete and precise stellar evolution theory to backtrack the history of each of these stars. 

\section*{Acknowledgment} 
H.F.X. acknowledges support from the Scientific and Technological Innovation Programs of Higher Education Institutions in Shanxi (STIP) (No. 2020L0528) and the Applied Basic Research Programs of Natural Science Foundation of Shanxi Province (No. 202103021223320).
J.S.N. acknowledges support from the National Natural Science Foundation of China (NSFC) (No. 12005124 and No. 12147215). 
J.N.F. acknowledges support from the National Natural Science Foundation of China (NSFC) (No. 11833002, No. 12090040, and No. 12090042). 

\software{{MESA} \citep{Paxton2011,Paxton2013,Paxton2015,Paxton2018,Paxton2019}, {GYRE} \citep{Townsend2013,Townsend2018,Goldstein2020}, {\tt emcee} \citep{emcee}}

\clearpage


\clearpage
\appendix

\section{Physical Parameters}
\label{sec:appA}
The physical parameters (e.g., the effective temperature $\Teff$, the effective surface gravity $\logg$ and the luminosity) show variations in a pulsation cycle of AE UMa. The effective temperature $\Teff$ varies from 7150 K to 8320 K, and the effective surface gravity $\logg$ varies from 3.77 to 4.16 \citep{Rodriguez1992,Hintz1997,Niu2017}.

About the luminosity, we calculated it based on the distance, apparent magnitude, extinction, and bolometric correction. The absolute magnitude $M_{V}$ can be obtained following the formula
\begin{equation}
   M_V=V-5\log{d}+5-A_V
   \label{eq:Mv}
\end{equation}
where $V=11.342$ mag was taken from AAVSO Photometric All Sky Survey (APASS) catalog \citep{Henden2016}, the distance $d=711$ pc was provided by Gaia DR3 \citep{Bailer-Jones2021}, the extinction $A_V=0.041$ mag was obtained from the maps of \citet{Schlafly2011}, and the lightness variation in $V$ band was about 0.6 mag \citep{Niu2017}.

On the other hand, the absolute bolometric magnitude $M_{bol}$ can be calculated from 
\begin{equation}
   M_{bol}=M_V+BC,
   \label{eq:Mbol}
\end{equation}
where the empirical bolometric correction
\begin{equation}
   BC=0.128\log{P}+0.022
   \label{eq:BC}
\end{equation}
for $\delta$ Scuti stars, which was derived by \citet{Petersen1999}.

Then, the luminosity can be obtained via 
\begin{equation}
   \log{L/ L_{\odot}} = -0.4 (M_{\rm{bol}} - M_{\rm{bol,\odot}})
   \label{eq:L}
\end{equation}
where the bolometric magnitude of the Sun $M_{\rm{bol,\odot}}=4.73$ mag was taken from Ref.  \citep{Torres2010}. Finally, we got the range of the observed luminosity as $\log{L/L_{\odot}}\in[1.00,1.24]$.

\clearpage

\section{Long Tables}
\label{sec:appB}

\startlongtable
\begin{deluxetable*}{ccccccc}
\centering
\tablecaption{Multi-frequency solution of the \TESS Sector 21 light curves of AE UMa. Note: $\sigma_f$ denotes the error estimation of frequency, $\sigma_a$ denotes the error estimation of amplitude. S/N is calculated within a spectral window of 2 $\cd$ equally divided by the frequency peak.  \label{tab:fre_solu}}
\tablehead{
\colhead{NO.}&\colhead{Marks}&\colhead{Frequency\ $(\rm{c\ days^{-1}})$} & \colhead{$\sigma_f$\ $(\rm{c\ days^{-1}})$} & \colhead{Amplitude\ (mmag)}&\colhead{$\sigma_a$\ (mmag)} &\colhead{S/N}
}
\startdata
F1& $f_0$	    &11.6257&  0.0002&	139.81& 1.69&  82.57 \\
F2& $2f_0$	    &23.2514&  0.0003&   48.90& 0.63&  77.80 \\
F3& $f_1$	    &15.0314&  0.0002&	 30.63& 0.38&  81.60 \\
F4& $f_0+f_1$	&26.6571&  0.0003&   20.63& 0.26&  79.42 \\
F5& $-f_0+f_1$	& 3.4057&  0.0003&   18.70& 0.23&  80.47 \\
F6& $3f_0$	    &34.8768&  0.0003&   18.38& 0.23&  78.37 \\
F7& $2f_0+f_1$	&38.2825&  0.0003&   11.21& 0.14&  79.01 \\
F8& $4f_0$	    &46.5024&  0.0003&    8.39& 0.12&  70.07 \\
F9& $3f_0+f_1$	&49.9079&  0.0003&    6.15& 0.08&  73.10 \\
F10&$2f_0-f_1$  & 8.2204&  0.0003&    5.13& 0.07&  68.95 \\
F11&$5f_0$	    &58.1277&  0.0004&    3.81& 0.07&  57.40 \\
F12&$4f_0+f_1$	&61.5335&  0.0003&    3.28& 0.06&  57.67 \\
F13&$f_0+2f_1$	&41.6886&  0.0004&    2.99& 0.06&  50.31\\
F14&$2f_1$	    &30.0629&  0.0004&    2.98& 0.06&  46.91 \\
F15&$5f_0+f_1$	&73.1590&  0.0004&    2.23& 0.04&  52.41\\
F16&$3f_0-f_1$	&19.8457&  0.0005&    2.17& 0.05&  39.56 \\
F17&$6f_0$  	&69.7532&  0.0005&    2.06& 0.04&  46.76 \\
F18&$2f_0+2f_1$ &53.3136&  0.0005&    1.63& 0.04&  36.90\\
F19&$6f_0+f_1$	&84.7849&  0.0005&    1.52& 0.04&  38.33\\
F20&$-f_0+2f_1$ &18.4368&  0.0005&    1.45& 0.04&  38.83\\
F21&$7f_0$      &81.3788&  0.0006&    1.21& 0.04&  32.91\\
F22&$3f_0+2f_1$ &64.9389&  0.0006&    1.12& 0.04&  31.57\\
F23&$7f_0+f_1$	&96.4101&  0.0006&    1.01& 0.03&  31.60\\
F24&$4f_0-f_1$	&31.4717&  0.0006&    1.01& 0.03&  31.53\\
F25&$4f_0+2f_1$ &76.5643&  0.0008&    0.87& 0.03&  26.28\\
F26&$-2f_0+2f_1$& 6.8110&  0.0009&    0.78& 0.04&  21.30\\
F27&$5f_0+2f_1$ &88.1900&  0.0010&    0.76& 0.04&  20.98\\
F28&$3f_0-2f_1$ & 4.8153&  0.0008&    0.74& 0.03&  25.00\\
F29&$8f_0$	    &93.0044&  0.0009&    0.72& 0.03&  23.00\\
F30&$8f_0+f_1$  &108.0342& 0.0009&    0.65& 0.03&  21.70\\
F31&$5f_0-f_1$	&43.0957&  0.0009&    0.63& 0.03&  21.62\\
F32&$6f_0-f_1$	&54.7205&  0.0011&    0.61& 0.03&  18.52\\
F33&$6f_0+2f_1$ &99.8155&  0.0010&    0.58& 0.03&  20.54\\
F34&$7f_0+2f_1$ &111.4425& 0.0011&    0.52& 0.03&  18.94\\
F35&$7f_0-f_1$	&66.3453&  0.0010&    0.52& 0.03&  20.33\\
F36&$f_0+3f_1$  &56.7191&  0.0014&    0.44& 0.03&  14.72\\
F37&$4f_0-2f_1$ &16.4431&  0.0013&    0.44& 0.03&  15.81\\
F38&$9f_0+f_1$  &119.6604& 0.0015&    0.43& 0.03&  13.75\\
F39&$3f_1$      &45.0948&  0.0013&    0.41& 0.03&  15.31\\
F40&$8f_0-f_1$	&77.9740&  0.0016&    0.38& 0.03&  12.39\\
F41&$9f_0$      &104.6287& 0.0017&    0.37& 0.03&  11.77\\
F42&$8f_0+2f_1$ &123.0679& 0.0016&    0.34& 0.03&  12.73\\
F43&$5f_0-2f_1$ &28.0635&  0.0017&    0.32& 0.03&  12.11\\
F44&$10f_0+f_1$ &131.2863& 0.0018&    0.30& 0.03&  11.02\\
F45&$9f_0+2f_1$ &134.6940& 0.0020&    0.29& 0.03&  10.02\\
F46&$9f_0-f_1$	&89.5975&  0.0023&    0.24& 0.03&	8.84\\
F47&$10f_0$     &116.2546& 0.0027&    0.24& 0.03&	7.58\\
F48&$2f_0+3f_1$ &68.3446&  0.0020&    0.24& 0.02&	9.98\\
F49&$3f_0+3f_1$ &79.9684&  0.0025&    0.22& 0.03&	8.10\\
F50&$4f_0+3f_1$ &91.5941&  0.0027&    0.21& 0.03&	7.57\\
F51&$5f_0+3f_1$ &103.2217& 0.0027&    0.21& 0.03&	7.37\\
F52&$10f_0+2f_1$&146.3188& 0.0027&    0.19& 0.03&   7.42\\
F53&$11f_0+f_1$ &142.9116& 0.0030&    0.19& 0.03&	6.86\\
F54&$7f_0+3f_1$ &126.4749& 0.0027&    0.18& 0.02&	7.47\\
F55&$6f_0-2f_1$ &39.6944&  0.0031&    0.18& 0.03&	6.42\\
F56&$6f_0+3f_1$ &114.8437& 0.0033&    0.18& 0.03&	6.11\\
F57&$8f_0+3f_1$ &138.1003& 0.0034&    0.17& 0.03&   5.91\\
F58&$-f_0+3f_1$ &33.4703&  0.0037&    0.16& 0.03&   5.49\\
F59&$10f_0-f_1$ &101.2166& 0.0039&    0.14& 0.03&   5.12\\
F60&$-2f_0+3f_1$&21.8381&  0.0043&    0.13& 0.03&	4.74\\
F61&$12f_0+f_1$ &154.5414& 0.0043&    0.13& 0.03&	4.71\\
F62&$5f_0-3f_1$ &13.0325&  0.0042&    0.12& 0.03&	4.77\\
F63&$11f_0$     &127.8854& 0.0042&    0.12& 0.02&	4.78\\
F64&$9f_0+3f_1$ &149.7320& 0.0050&    0.12& 0.03&   4.06\\
F65&$7f_0-2f_1$ &51.3246&  0.0047&    0.11& 0.03&   4.28\\
\enddata
\end{deluxetable*}

\clearpage

\startlongtable
\begin{deluxetable*}{ccc|ccc|ccc}
  \centering
  \tablecaption{All times of maximum light used in this work. $\sigma$: the uncertainties; Mark: the sources. \label{tab:O-C}}
  \tablehead{
  \colhead{$\rm{BJD_{TDB}}$}&\colhead{$\sigma$}&\colhead{Mark}&\colhead{$\rm{BJD_{TDB}}$}&\colhead{$\sigma$}&\colhead{Mark}&\colhead{$\rm{BJD_{TDB}}$}&\colhead{$\sigma$}&\colhead{Mark}
  }
  \startdata
2442062.5837  &0.0001  &1   &2454791.9945  &0.0005  &14  &2457466.4354 	    &0.0006   &25	 \\
2442065.5964  &0.0001  &2   &2454807.8229  &0.0006  &14  &2457712.78688 	&0.00002  &26	 \\
2442065.6783  &0.0001  &2   &2454807.9039  &0.0005  &14  &2457712.87055 	&0.00003  &26	 \\
2442068.3437  &0.0001  &2   &2454807.9873  &0.0007  &14  &2457712.96264 	&0.00006  &26	 \\
2442068.4307  &0.0001  &2   &2454816.6800  &0.0006  &14  &2457736.69950 	&0.00008  &26	 \\
2442068.5208  &0.0001  &2   &2454816.7619  &0.0006  &14  &2457736.79017 	&0.00003  &26	 \\
2442068.6034  &0.0001  &2   &2454816.8508  &0.0012  &14  &2457737.73234 	&0.00002  &26	 \\
2442068.6876  &0.0001  &2   &2454816.9406  &0.0009  &14  &2457737.81526 	&0.00002  &26	 \\
2442069.3813  &0.0001  &2   &2454821.83920 &0.00012 &26  &2457750.71793 	&0.00002  &26	 \\
2442069.4656  &0.0001  &2   &2454821.92866 &0.00010 &26  &2457754.68162 	&0.00004  &26	 \\
2442069.5478  &0.0001  &2   &2454823.81476 &0.00007 &26  &2457754.76328 	&0.00001  &26	 \\
2442069.6368  &0.0001  &2   &2454823.90464 &0.00011 &26  &2457754.84626 	&0.00002  &26	 \\
2442086.4970  &0.0001  &2   &2454828.89354 &0.00012 &26  &2457754.93784 	&0.00003  &26	 \\
2442086.5792  &0.0001  &2   &2454831.81665 &0.00004 &26  &2457759.75330 	&0.00001  &26	 \\
2442087.4395  &0.0001  &2   &2454831.90727 &0.00006 &26  &2457759.92720 	&0.00005  &26	 \\
2442087.5268  &0.0001  &2   &2454836.80564 &0.00005 &26  &2457760.01375 	&0.00002  &26	 \\
2442087.6160  &0.0001  &2   &2454836.89619 &0.00006 &26  &2457760.61471 	&0.00002  &26	 \\
2442095.5303  &0.0001  &1   &2454837.6644  &0.0009  &14  &2457760.69625 	&0.00002  &26	 \\
2442095.6128  &0.0001  &1   &2454837.7572  &0.0009  &14  &2457760.78462 	&0.00003  &26	 \\
2442103.3518  &0.0001  &2   &2454837.83872 &0.00004 &26  &2457760.87458 	&0.00002  &26	 \\
2442106.4528  &0.0001  &1   &2454840.68081 &0.00009 &26  &2457760.95547 	&0.00001  &26	 \\
2442119.5257  &0.0001  &1   &2454840.76418 &0.00003 &26  &2457761.73499 	&0.00005  &26	 \\
2442121.5030  &0.0001  &1   &2454840.84693 &0.00003 &26  &2457761.81728 	&0.00002  &26	 \\
2442122.3633  &0.0001  &2   &2454840.93604 &0.00009 &26  &2457761.90007 	&0.00002  &26	 \\
2442122.4489  &0.0001  &2   &2454843.6893  &0.0007  &14  &2457761.99147 	&0.00005  &26	 \\
2442128.2973  &0.0001  &1   &2454843.7711  &0.0008  &14  &2457799.58143 	&0.00011  &26	 \\
2442128.3877  &0.0001  &1   &2454843.8574  &0.0009  &14  &2457799.66695 	&0.00006  &26	 \\
2442128.4732  &0.0001  &1   &2454843.9501  &0.0010  &14  &2457852.64846 	&0.00002  &26	 \\
2442128.5562  &0.0001  &1   &2454846.6144  &0.0008  &14  &2457852.74036 	&0.00003  &26	 \\
2442133.4632  &0.0001  &1   &2454846.6967  &0.0007  &14  &2458181.40908 	&0.00001  &26	 \\
2442133.5447  &0.0001  &1   &2454847.7343  &0.0008  &14  &2458199.38347 	&0.00003  &26	 \\
2442134.4060  &0.0001  &1   &2454847.8171  &0.0007  &14  &2458229.40617 	&0.00007 &26	 \\
2442147.3938  &0.0001  &1   &2454847.8995  &0.0006  &14  &2458231.38243 	&0.00004 &26	 \\
2442148.4300  &0.0001  &1   &2454848.84885 &0.00007 &26  &2458531.49304 	&0.00006 &26	 \\
2442148.5122  &0.0001  &1   &2454848.93873 &0.00005 &26  &2458531.58339 	&0.00012 &26	 \\
2442159.4370  &0.0001  &1   &2454852.63050 &0.00005 &26  &2458539.58589 	&0.00007 &26	 \\
2442161.4150  &0.0001  &1   &2454852.72310 &0.00009 &26  &2458539.66642 	&0.00004 &26	 \\
2442453.5311  &0.0001  &1   &2454852.80649 &0.00006 &26  &2458857.93322 	&0.00002 &26	 \\
2442453.6142  &0.0001  &1   &2454852.88857 &0.00003 &26  &2458870.49564 	&0.00005 &26	 \\
2442460.4994  &0.0001  &1   &2454854.78810 &0.00006 &26  &2458870.57721 	&0.00002 &26	 \\
2442830.6285  &0.0001  &1   &2454854.87030 &0.00003 &26  &2458870.66063 	&0.00003 &26	 \\
2442837.5125  &0.0001  &1   &2454854.95318 &0.00003 &26  &2458870.75379 	&0.00003 &26	 \\
2442838.4596  &0.0001  &1   &2454855.6476  &0.0010  &14  &2458870.83771 	&0.00002 &26	 \\
2442869.3382  &0.0001  &1   &2454855.7323  &0.0010  &14  &2458870.91910 	&0.00002 &26	 \\
2442869.4210  &0.0001  &1   &2454855.8126  &0.0006  &14  &2458871.00836 	&0.00003 &26	 \\
2443162.5713  &0.0001  &1   &2454855.9030  &0.0010  &14  &2458871.09802 	&0.00002 &26	 \\
2444633.4631  &0.0001  &1   &2454862.70116 &0.00007 &26  &2458871.17798 	&0.00002 &26	 \\
2444633.5445  &0.0001  &1   &2454862.78468 &0.00002 &26  &2458871.26286 	&0.00006 &26	 \\
2444633.6314  &0.0001  &1   &2454862.86687 &0.00002 &26  &2458871.35589 	&0.00005 &26	 \\
2444634.4051  &0.0001  &1   &2454862.95571 &0.00008 &26  &2458871.43830 	&0.00002 &26	 \\
2444634.4907  &0.0001  &1   &2454864.5880  &0.0005  &14  &2458871.52069 	&0.00001 &26	 \\
2444634.5815  &0.0001  &1   &2454864.6731  &0.0007  &14  &2458871.61190 	&0.00003 &26	 \\
2444692.4714  &0.0001  &1   &2454864.7656  &0.0009  &14  &2458871.69903 	&0.00003 &26	 \\
2445355.4908  &0.0001  &1   &2454864.8484  &0.0006  &14  &2458871.77983 	&0.00001 &26	 \\
2445355.5733  &0.0001  &1   &2454864.9303  &0.0006  &14  &2458871.86637 	&0.00003 &26	 \\
2445382.3234  &0.0001  &1   &2454867.77368 &0.00003 &26  &2458871.95815 	&0.00004 &26	 \\
2445382.4110  &0.0001  &1   &2454867.85560 &0.00002 &26  &2458872.03962 	&0.00001 &26	 \\
2445382.5003  &0.0001  &1   &2454867.94563 &0.00006 &26  &2458872.12341 	&0.00002 &26	 \\
2445382.5813  &0.0001  &1   &2454868.6353  &0.0008  &14  &2458872.21532 	&0.00004 &26	 \\
2446468.4607  &0.0001  &1   &2454868.7173  &0.0008  &14  &2458872.30044 	&0.00001 &26	 \\
2446468.5474  &0.0001  &1   &2454868.80341 &0.00004 &26  &2458872.38140 	&0.00002 &26	 \\
2446855.6285  &0.0001  &3   &2454868.8040  &0.0014  &14  &2458872.47010 	&0.00002 &26	 \\
2446856.5735  &0.0001  &3   &2454868.8948  &0.0009  &14  &2458872.56038 	&0.00001 &26	 \\
2446856.6567  &0.0001  &3   &2454868.89492 &0.00006 &26  &2458872.64117 	&0.00001 &26	 \\
2446857.6023  &0.0001  &3   &2454869.83759 &0.00003 &26  &2458872.72584 	&0.00002 &26	 \\
2446857.6931  &0.0001  &3   &2454869.92022 &0.00004 &26  &2458872.81831 	&0.00003 &26	 \\
2446858.6388  &0.0001  &3   &2454871.81962 &0.00006 &26  &2458872.90151 	&0.00001 &26	 \\
2446859.6672  &0.0001  &3   &2454871.90141 &0.00004 &26  &2458872.98277 	&0.00002 &26	 \\
2446878.4187  &0.0001  &3   &2454877.41053 &0.00008 &26  &2458873.07347 	&0.00004 &26	 \\
2446878.5070  &0.0001  &3   &2454877.57609 &0.00006 &26  &2458873.16169 	&0.00002 &26	 \\
2446878.5952  &0.0001  &3   &2454878.6942  &0.0006  &14  &2458873.24229 	&0.00001 &26	 \\
2446884.5268  &0.0001  &3   &2454878.7818  &0.0008  &14  &2458873.32889 	&0.00002 &26	 \\
2446884.6123  &0.0001  &3   &2454878.8728  &0.0009  &14  &2458873.42074 	&0.00003 &26	 \\
2446886.5913  &0.0001  &3   &2454880.84619 &0.00005 &26  &2458873.50236 	&0.00002 &26	 \\
2450151.4571  &0.0001  &1   &2454881.36027 &0.00004 &26  &2458873.58521 	&0.00001 &26	 \\
2450151.5391  &0.0001  &1   &2454884.54483 &0.00004 &26  &2458873.67699 	&0.00004 &26	 \\
2450152.3177  &0.0001  &1   &2454889.71021 &0.00008 &26  &2458873.76298 	&0.00001 &26	 \\
2450152.4869  &0.0001  &1   &2454889.79400 &0.00003 &26  &2458873.84329 	&0.00002 &26	 \\
2450152.5763  &0.0001  &1   &2454889.87645 &0.00003 &26  &2458873.93138 	&0.00003 &26	 \\
2450458.8822  &0.0001  &4   &2454893.66178 &0.00003 &26  &2458874.02252 	&0.00003 &26	 \\
2450458.9643  &0.0001  &4   &2454894.4419  &0.0005  &17  &2458874.10328 	&0.00002 &26	 \\
2450459.8247  &0.0001  &4   &2454894.5234  &0.0005  &17  &2458874.18748 	&0.00001 &26	 \\
2450459.9120  &0.0001  &4   &2454894.6078  &0.0006  &17  &2458874.28015 	&0.00003 &26	 \\
2450467.7395  &0.0001  &4   &2454895.64489 &0.00004 &26  &2458874.36398 	&0.00001 &26	 \\
2450467.8243  &0.0001  &4   &2454895.72587 &0.00003 &26  &2458874.44476 	&0.00003 &26	 \\
2450490.3614  &0.0001  &1   &2454895.81257 &0.00007 &26  &2458874.53515 	&0.00003 &26	 \\
2450505.6704  &0.0001  &4   &2454895.90468 &0.00019 &26  &2458874.62426 	&0.00003 &26	 \\
2450505.7602  &0.0001  &4   &2454897.01955 &0.00009 &19  &2458874.70470 	&0.00001 &26	 \\
2450505.8468  &0.0001  &4   &2454897.10767 &0.00008 &19  &2458874.79012 	&0.00003 &26	 \\
2450516.7683  &0.0001  &4   &2454897.18907 &0.00004 &19  &2458874.88281 	&0.00004 &26	 \\
2450554.4439  &0.0001  &1   &2454897.36693 &0.00007 &26  &2458874.96478 	&0.00002 &26	 \\
2450813.3557  &0.0001  &1   &2454898.04976 &0.0010  &19  &2458875.04727 	&0.00001 &26	 \\
2450813.4415  &0.0001  &1   &2454898.13391 &0.0015  &19  &2458875.13891 	&0.00004 &26	 \\
2450813.6158  &0.0001  &1   &2454898.3091  &0.0004  &17  &2458875.22538 	&0.00002 &26	 \\
2450813.6992  &0.0001  &1   &2454898.99384 &0.00013 &19  &2458875.30597 	&0.00001 &26	 \\
2450848.4547  &0.0001  &1   &2454899.08448 &0.00023 &19  &2458875.39338 	&0.00003 &26	 \\
2450848.5398  &0.0001  &1   &2454899.17137 &0.00013 &19  &2458875.48491 	&0.00003 &26	 \\
2450848.6219  &0.0001  &1   &2454899.59626 &0.00006 &26  &2458875.56600 	&0.00002 &26	 \\
2450849.4822  &0.0001  &1   &2454899.68821 &0.00008 &26  &2458875.64951 	&0.00001 &26	 \\
2450849.5695  &0.0001  &1   &2454899.77260 &0.00004 &26  &2458875.74197 	&0.00005 &26	 \\
2450862.3847  &0.0001  &5   &2454899.85460 &0.00004 &26  &2458875.82662 	&0.00001 &26	 \\
2450862.38719 &0.00011 &26  &2454900.03261 &0.00010 &19  &2458875.90783 	&0.00002 &26	 \\
2450872.2816  &0.0001  &1   &2454900.11364 &0.0008  &19  &2458875.99683 	&0.00005 &26	 \\
2450872.3641  &0.0001  &1   &2454900.19858 &0.00014 &19  &2458876.08681 	&0.00003 &26	 \\
2450872.4488  &0.0001  &1   &2454901.05834 &0.00027 &19  &2458876.16728 	&0.00001 &26	 \\
2450872.5401  &0.0001  &1   &2454902.77954 &0.00003 &26  &2458876.25200 	&0.00003 &26	 \\
2450899.3736  &0.0001  &1   &2454904.4206  &0.0007  &17  &2458876.34479 	&0.00004 &26	 \\
2450899.4577  &0.0001  &1   &2454904.5013  &0.0004  &17  &2458876.42775 	&0.00002 &26	 \\
2450902.2983  &0.0001  &1   &2454907.34502 &0.00008 &26  &2458876.50956 	&0.00002 &26	 \\
2450902.3826  &0.0001  &1   &2454907.42676 &0.00005 &26  &2458876.60048 	&0.00005 &26	 \\
2450903.3328  &0.0001  &1   &2454909.3154  &0.0004  &17  &2458876.68821 	&0.00003 &26	 \\
2450903.4199  &0.0001  &1   &2454909.4094  &0.0006  &17  &2458876.76860 	&0.00002 &26	 \\
2450903.5016  &0.0001  &1   &2454909.4902  &0.0006  &17  &2458876.85513 	&0.00002 &26	 \\
2451269.509   &0.002   &6   &2454910.4342  &0.0008  &17  &2458876.94718 	&0.00003 &26	 \\
2451283.442   &0.002   &6   &2454912.3330  &0.0012  &17  &2458877.02860 	&0.00002 &26	 \\
2451283.526   &0.002   &6   &2454912.4153  &0.0004  &17  &2458877.11156 	&0.00002 &26	 \\
2451283.610   &0.001   &6   &2454924.3749  &0.0040  &17  &2458877.20379 	&0.00005 &26	 \\
2451318.364   &0.001   &6   &2454926.60542 &0.00006 &26  &2458877.28943 	&0.000009&26	 \\
2451318.447   &0.002   &6   &2454926.69757 &0.00013 &26  &2458877.36994 	&0.00002 &26	 \\
2451608.07238 &0.00001 &7   &2454944.58674 &0.00007 &26  &2458877.45800 	&0.00003 &26	 \\
2451608.15849 &0.00001 &7   &2454944.67459 &0.00005 &26  &2458877.54912 	&0.00004 &26	 \\
2451608.24027 &0.00001 &7   &2454946.65202 &0.00007 &26  &2458877.62968 	&0.00002 &26	 \\
2451608.32715 &0.00001 &7   &2454947.59993 &0.00005 &26  &2458877.71348 &0.00004 &26	 \\
2451609.01933 &0.00001 &7   &2454947.68109 &0.00003 &26  &2458877.80691 	&0.00005 &26	 \\
2451609.10137 &0.00001 &7   &2454947.76600 &0.00009 &26  &2458877.89003 	&0.00004 &26	 \\
2451609.18725 &0.00001 &7   &2454949.57565 &0.00007 &26  &2458877.97193 	&0.00002 &26	 \\
2451609.27776 &0.00001 &7   &2454949.66396 &0.00003 &26  &2458878.06198 	&0.00006 &26	 \\
2451609.35903 &0.00001 &7   &2454951.64071 &0.00006 &26  &2458878.15057 	&0.00002 &26	 \\
2451610.04573 &0.00001 &7   &2454951.72793 &0.00004 &26  &2458878.23097 	&0.00002 &26	 \\
2451610.99764 &0.00001 &7   &2454970.04393 &0.00008 &19  &2458878.31738 	&0.00003 &26	 \\
2451611.08286 &0.00001 &7   &2454972.10855 &0.00024 &19  &2458878.40948 	&0.00003 &26	 \\
2451611.16345 &0.00001 &7   &2455199.71023 &0.00009 &26  &2458878.49120 	&0.00001 &26	 \\
2451612.02532 &0.00001 &7   &2455199.80216 &0.00014 &26  &2458878.57373 	&0.00002 &26	 \\
2451612.10980 &0.00001 &7   &2455199.88467 &0.00006 &26  &2458878.66541 	&0.00006 &26	 \\
2451612.20179 &0.00001 &7   &2455202.63391 &0.00008 &26  &2458878.75171 	&0.00002 &26	 \\
2451612.28538 &0.00001 &7   &2455202.72659 &0.00015 &26  &2458878.83250 	&0.00001 &26	 \\
2451612.37118 &0.00001 &7   &2455202.80999 &0.00007 &26  &2458878.92007 	&0.00005 &26	 \\
2451612.96996 &0.00001 &7   &2455223.44980 &0.00004 &26  &2458879.01134 	&0.00003 &26	 \\
2451613.06164 &0.00001 &7   &2455223.54053 &0.00006 &26  &2458879.09225 	&0.00003 &26	 \\
2451613.14610 &0.00001 &7   &2455223.62840 &0.00005 &26  &2458879.17607 	&0.00003 &26	 \\
2451613.22839 &0.00001 &7   &2455230.24632 &0.00009 &19  &2458879.26871 	&0.00006 &26	 \\
2451613.31631 &0.00001 &7   &2455230.33867 &0.00010 &19  &2458879.35296 	&0.00003 &26	 \\
2451615.03489 &0.00001 &7   &2455230.42086 &0.00008 &19  &2458879.43417 	&0.00002 &26	 \\
2451615.12679 &0.00001 &7   &2455231.28257 &0.00008 &19  &2458879.52368 	&0.00004 &26	 \\
2451615.21055 &0.00001 &7   &2455231.36431 &0.00006 &19  &2458879.61314 	&0.00002 &26	 \\
2451615.29269 &0.00001 &7   &2455233.34646 &0.00003 &26  &2458879.69355 	&0.00002 &26	 \\
2451615.98622 &0.00001 &7   &2455233.42847 &0.00002 &26  &2458879.77905 	&0.00002 &26	 \\
2451616.07122 &0.00001 &7   &2455233.51836 &0.00004 &26  &2458879.87182 	&0.00004 &26	 \\
2451616.15338 &0.00001 &7   &2455233.60682 &0.00003 &26  &2458879.95373 	&0.00003 &26	 \\
2451616.23960 &0.00001 &7   &2455238.76197 &0.00005 &26  &2458880.03596 	&0.00002 &26	 \\
2451929.25633 &0.00001 &7   &2455244.44561 &0.00004 &26  &2458880.12706 	&0.00008 &26	 \\
2451929.34714 &0.00001 &7   &2455244.52721 &0.00002 &26  &2458880.21440 	&0.00003 &26	 \\
2451930.20658 &0.00001 &7   &2455244.61083 &0.00003 &26  &2458880.29527 	&0.00002 &26	 \\
2451930.28924 &0.00001 &7   &2455244.70247 &0.00007 &26  &2458880.38204 	&0.00003 &26	 \\
2451930.37279 &0.00001 &7   &2455257.69079 &0.00004 &26  &2458880.47386 	&0.00003 &26	 \\
2451931.23228 &0.00001 &7   &2455257.77210 &0.00002 &26  &2458880.55507 	&0.00002 &26	 \\
2451931.32102 &0.00001 &7   &2455259.4115  &0.0014  &18  &2458880.63841 	&0.00001 &26	 \\
2451941.21094 &0.00001 &7   &2455259.66868 &0.00009 &26  &2458880.73072 	&0.00004 &26	 \\
2451941.29867 &0.00001 &7   &2455259.75465 &0.00005 &26  &2458880.81559 	&0.00002 &26	 \\
2451941.38884 &0.00001 &7   &2455266.46539 &0.00006 &26  &2458880.89664 	&0.00002 &26	 \\
2451942.15698 &0.00001 &7   &2455266.54739 &0.00005 &26  &2458880.98571 	&0.00013 &26	 \\
2451942.24803 &0.00001 &7   &2455266.63087 &0.00007 &26  &2458881.07546 	&0.00005 &26	 \\
2451942.33180 &0.00001 &7   &2455279.62366 &0.00010 &26  &2458881.15606 	&0.00003 &26	 \\
2451942.41487 &0.00001 &7   &2455279.71180 &0.00007 &26  &2458881.24108 	&0.00003 &26	 \\
2452321.4096  &0.0008  &8   &2455282.63677 &0.00005 &26  &2458881.33315 	&0.00006 &26	 \\
2452371.3843  &0.0002  &8   &2455282.71747 &0.00003 &26  &2458881.41665 	&0.00002 &26	 \\
2452685.3467  &0.0012  &8   &2455293.3833  &0.0010  &18  &2458881.49844 	&0.00003 &26	 \\
2452685.4376  &0.0012  &8   &2455293.4759  &0.0007  &18  &2458881.58896 	&0.00006 &26	 \\
2452730.4194  &0.0004  &8   &2455298.62926 &0.00003 &26  &2458881.67686 	&0.00003 &26	 \\
2452739.3624  &0.0007  &8   &2455301.64383 &0.00007 &26  &2458881.75726 	&0.00004 &26	 \\
2452745.4709  &0.0004  &9   &2455301.73217 &0.00005 &26  &2458881.84352 	&0.00003 &26	 \\
2452980.5615  &0.0011  &10  &2455302.3325  &0.0009  &18  &2458881.93608 	&0.00003 &26	 \\
2452980.6428  &0.0011  &10  &2455303.3598  &0.0014  &18  &2458882.01766 	&0.00002 &26	 \\
2452980.7286  &0.0011  &10  &2455304.3966  &0.0007  &18  &2458882.10057 	&0.00003 &26	 \\
2453003.5238  &0.0004  &9   &2455304.4782  &0.0010  &18  &2458882.19256 	&0.00005 &26	 \\
2453028.2949  &0.0006  &9   &2455305.3395  &0.0007  &18  &2458882.27817 	&0.00002 &26	 \\
2453028.3878  &0.0010  &9   &2455305.4245  &0.0014  &18  &2458882.35909 	&0.00002 &26	 \\
2453028.4712  &0.0005  &9   &2455309.3855  &0.0008  &18  &2458882.44663 	&0.00006 &26	 \\
2453028.5529  &0.0006  &9   &2455310.4131  &0.0019  &18  &2458882.53785 	&0.00004 &26	 \\
2453028.6427  &0.0011  &9   &2455311.3669  &0.0007  &18  &2458882.61895 	&0.00001 &26	 \\
2453069.4126  &0.0016  &9   &2455311.4495  &0.0005  &18  &2458882.70329 	&0.00003 &26	 \\
2453069.5036  &0.0018  &9   &2455314.63560 &0.00008 &26  &2458882.79554 	&0.00007 &26	 \\
2453070.4500  &0.0013  &9   &2455554.87787 &0.00003 &26  &2458882.87923 	&0.00003 &26	 \\
2453070.533   &0.002   &9   &2455554.96003 &0.00003 &26  &2458882.96052 	&0.00004 &26	 \\
2453090.4060  &0.0027  &9   &2455555.73931 &0.00003 &26  &2458883.05111 	&0.00008 &26	 \\
2453094.3582  &0.0010  &9   &2455555.82099 &0.00002 &26  &2458883.13989 	&0.00002 &26	 \\
2453110.3626  &0.0019  &12  &2455555.90869 &0.00007 &26  &2458883.21988 	&0.00002 &26	 \\
2453409.52929 &0.00024 &11  &2455562.79234 &0.00004 &26  &2458883.30536 	&0.00006 &26	 \\
2453409.61157 &0.00032 &11  &2455562.87351 &0.00004 &26  &2458883.39844 	&0.00009 &26	 \\
2453409.69477 &0.00058 &11  &2455562.96285 &0.00011 &26  &2458883.48044 	&0.00002 &26	 \\
2453427.3279  &0.0011  &12  &2455583.43210 &0.00012 &19  &2458883.56278 	&0.00005 &26	 \\
2453427.4188  &0.0012  &12  &2455584.37947 &0.00018 &19  &2458883.65416 	&0.00003 &26	 \\
2453427.5038  &0.0012  &12  &2455584.55239 &0.00008 &26  &2458883.74089 	&0.00002 &26	 \\
2453451.4143  &0.0022  &10  &2455584.63618 &0.00008 &26  &2458883.82168 	&0.00001 &26	 \\
2453451.5049  &0.0024  &10  &2455584.72789 &0.00014 &26  &2458883.90864 	&0.00003 &26	 \\
2453484.3623  &0.0024  &10  &2455584.81253 &0.00006 &26  &2458886.74639 	&0.00002 &26	 \\
2453765.3810  &0.0002  &13  &2455584.89442 &0.00006 &26  &2458886.83222 	&0.00004 &26	 \\
2453765.4667  &0.0002  &13  &2455584.98193 &0.00019 &26  &2458886.92482 	&0.00007 &26	 \\
2453765.5469  &0.0002  &13  &2455585.41399 &0.00008 &19  &2458886.92486 	&0.00007 &26	 \\
2453766.3285  &0.0002  &13  &2455587.39535 &0.00009 &19  &2458887.00676 	&0.00003 &26	 \\
2453766.4086  &0.0002  &13  &2455588.42086 &0.00008 &19  &2458887.08949 	&0.00003 &26	 \\
2453766.4950  &0.0002  &13  &2455607.95360 &0.00018 &19  &2458887.18135 	&0.00003 &26	 \\
2453766.5856  &0.0002  &13  &2455608.03562 &0.00011 &19  &2458887.26740 	&0.00003 &26	 \\
2453794.3626  &0.0003  &11  &2455608.11878 &0.00011 &19  &2458887.34802 	&0.00002 &26	 \\
2453795.3122  &0.0003  &11  &2455608.21064 &0.00022 &19  &2458887.43575 	&0.00004 &26	 \\
2453795.4005  &0.0001  &11  &2455608.29613 &0.00012 &19  &2458887.52696 	&0.00004 &26	 \\
2453827.6577  &0.0013  &11  &2455608.37710 &0.00011 &19  &2458887.60792 	&0.00002 &26	 \\
2453827.65791 &0.00014 &26  &2455610.35990 &0.00010 &19  &2458887.69180 	&0.00003 &26	 \\
2454037.88265 &0.00013 &26  &2455611.38852 &0.00016 &19  &2458887.78471 	&0.00005 &26	 \\
2454050.87089 &0.00013 &26  &2455612.24770 &0.00014 &19  &2458887.86849 	&0.00002 &26	 \\
2454050.95159 &0.00010 &26  &2455612.33995 &0.00022 &19  &2458887.94978 	&0.00002 &26	 \\
2454056.80189 &0.00008 &26  &2455613.19877 &0.00021 &19  &2458888.03946 	&0.00009 &26	 \\
2454056.88685 &0.00007 &26  &2455616.98081 &0.00011 &19  &2458888.12864 	&0.00003 &26	 \\
2454079.7673  &0.0007  &14  &2455617.07131 &0.00008 &19  &2458888.20918 	&0.00002 &26	 \\
2454079.8587  &0.0014  &14  &2455617.15276 &0.00005 &19  &2458888.29435 	&0.00004 &26	 \\
2454079.9444  &0.0007  &14  &2455617.23661 &0.00006 &19  &2458888.38707 	&0.00005 &26	 \\
2454087.77092 &0.00012 &26  &2455617.32888 &0.00010 &19  &2458888.46953 	&0.00002 &26	 \\
2454102.6478  &0.0006  &14  &2455620.07795 &0.00010 &19  &2458888.55188 	&0.00001 &26	 \\
2454102.7372  &0.0015  &14  &2455620.16102 &0.00017 &19  &2458888.64285 	&0.00005 &26	 \\
2454102.8261  &0.0009  &14  &2455620.33798 &0.00007 &26  &2458888.72998 	&0.00002 &26	 \\
2454103.6856  &0.0009  &14  &2455621.11110 &0.00017 &19  &2458888.81072 	&0.00002 &26	 \\
2454103.7684  &0.0012  &14  &2455621.19938 &0.00019 &19  &2458888.89759 	&0.00003 &26	 \\
2454104.7124  &0.0009  &14  &2455878.81472 &0.00003 &26  &2458888.98939 	&0.00002 &26	 \\
2454107.6365  &0.0005  &14  &2455878.90285 &0.00005 &26  &2458889.07032 	&0.00003 &26	 \\
2454107.7258  &0.0009  &14  &2455890.77593 &0.00008 &26  &2458889.15393 	&0.00002 &26	 \\
2454107.8145  &0.0010  &14  &2455890.85714 &0.00007 &26  &2458889.24631 	&0.00004 &26	 \\
2454107.8951  &0.0007  &14  &2455890.94564 &0.00012 &26  &2458889.33116 	&0.00002 &26	 \\
2454110.7395  &0.0011  &14  &2455907.72203 &0.00012 &26  &2458889.41215 	&0.00001 &26	 \\
2454110.8214  &0.0006  &14  &2455907.80705 &0.00007 &26  &2458889.50077 	&0.00006 &26	 \\
2454110.90530 &0.00009 &26  &2455931.71546 &0.00010 &26  &2458889.59125 	&0.00003 &26	 \\
2454110.9058  &0.0007  &14  &2455931.80728 &0.00016 &26  &2458889.67192 	&0.00002 &26	 \\
2454110.9984  &0.0007  &14  &2455937.39891 &0.00012 &19  &2458889.75649 	&0.00002 &26	 \\
2454115.72933 &0.00013 &26  &2455961.30584 &0.00004 &19  &2458889.84925 	&0.00004 &26	 \\
2454121.66129 &0.00009 &26  &2455961.39147 &0.00010 &19  &2458889.93229 	&0.00002 &26	 \\
2454121.74340 &0.00009 &26  &2455964.31556 &0.00014 &19  &2458890.01399 	&0.00002 &26	 \\
2454125.7063  &0.0010  &14  &2455964.40805 &0.00019 &19  &2458890.10432 	&0.00007 &26	 \\
2454125.7890  &0.0011  &14  &2455966.21408 &0.00007 &19  &2458890.19276 	&0.00002 &26	 \\
2454125.8723  &0.0008  &14  &2455966.29493 &0.00005 &19  &2458890.27331 	&0.00002 &26	 \\
2454125.9656  &0.0014  &14  &2455966.38042 &0.00008 &19  &2458890.35946 	&0.00005 &26	 \\
2454131.5557  &0.0013  &14  &2455967.33183 &0.00009 &19  &2458890.45191 	&0.00006 &26	 \\
2454131.6399  &0.0008  &14  &2455967.41669 &0.00007 &19  &2458890.53334 	&0.00002 &26	 \\
2454131.7219  &0.0006  &14  &2455968.27811 &0.00007 &19  &2458890.61635 	&0.00002 &26	 \\
2454131.8113  &0.0011  &14  &2455968.35894 &0.00004 &19  &2458890.70829 	&0.00005 &26	 \\
2454131.8996  &0.0006  &14  &2455969.30458 &0.00007 &19  &2458890.79389 	&0.00001 &26	 \\
2454136.5453  &0.0011  &14  &2455969.39671 &0.00014 &19  &2458890.87450 	&0.00003 &26	 \\
2454136.6282  &0.0006  &14  &2455984.4479  &0.0014  &20  &2458890.96239 	&0.00005 &26	 \\
2454136.7111  &0.0008  &14  &2455993.30430 &0.0005  &22  &2458891.05366 	&0.00003 &26	 \\
2454136.8005  &0.0009  &14  &2455993.38780 &0.0005  &22  &2458891.13421 	&0.00002 &26	 \\
2454136.8889  &0.0007  &14  &2455997.00877 &0.00008 &19  &2458891.21844 	&0.00003 &26	 \\
2454136.9704  &0.0012  &14  &2455997.09141 &0.00004 &19  &2458891.31115 	&0.00005 &26	 \\
2454137.6588  &0.0013  &14  &2455997.17372 &0.00004 &19  &2458891.39493 	&0.00001 &26	 \\
2454137.7498  &0.0014  &14  &2455997.26468 &0.00010 &19  &2458891.56607 	&0.00009 &26	 \\
2454137.8318  &0.0009  &14  &2455998.03450 &0.00008 &19  &2458891.65506 	&0.00004 &26	 \\
2454139.7233  &0.0009  &14  &2455998.12286 &0.00014 &19  &2458891.73556 	&0.00002 &26	 \\
2454139.8131  &0.0008  &14  &2455998.21269 &0.00008 &19  &2458891.82131 	&0.00003 &26	 \\
2454139.8949  &0.0005  &14  &2455998.29351 &0.00006 &19  &2458891.91382 	&0.00005 &26	 \\
2454141.7883  &0.0010  &14  &2455998.98155 &0.00009 &19  &2458891.99596 	&0.00002 &26	 \\
2454141.8775  &0.0007  &14  &2455999.07344 &0.00009 &19  &2458892.07860 	&0.00001 &26	 \\
2454141.9582  &0.0009  &14  &2455999.15543 &0.00005 &19  &2458892.17001 	&0.00005 &26	 \\
2454151.67767 &0.00009 &26  &2455999.23823 &0.00005 &19  &2458892.25648 	&0.00003 &26	 \\
2454151.76589 &0.00027 &26  &2456000.01689 &0.00005 &19  &2458892.33707 	&0.00002 &26	 \\
2454152.6248  &0.0011  &14  &2456000.09857 &0.00004 &19  &2458892.42443 	&0.00003 &26	 \\
2454153.5753  &0.0010  &14  &2456000.18819 &0.00009 &19  &2458892.51619 	&0.00004 &26	 \\
2454154.6890  &0.0010  &14  &2456000.27695 &0.00006 &19  &2458892.59699 	&0.00002 &26	 \\
2454154.7807  &0.0009  &14  &2456001.04671 &0.00006 &19  &2458892.68062 	&0.00002 &26	 \\
2454154.8625  &0.0006  &14  &2456001.13759 &0.00006 &19  &2458892.77309 	&0.00006 &26	 \\
2454164.66673 &0.00021 &26  &2456001.21934 &0.00003 &19  &2458892.85766 	&0.00001 &26	 \\
2454164.75938 &0.00022 &26  &2456001.30305 &0.00006 &19  &2458892.93896 	&0.00001 &26	 \\
2454171.4671  &0.0023  &15  &2456002.4214  &0.0006  &23  &2458893.02790 	&0.00007 &26	 \\
2454175.4213  &0.0013  &15  &2456006.4683  &0.0001  &23  &2458893.11768 	&0.00004 &26	 \\
2454184.62265 &0.00008 &26  &2456008.3559  &0.0008  &23  &2458893.19803 	&0.00004 &26	 \\
2454184.71548 &0.00019 &26  &2456008.4486  &0.0012  &23  &2458893.28354 	&0.00002 &26	 \\
2454188.58406 &0.00010 &26  &2456009.4745  &0.0007  &23  &2458893.37609 	&0.00005 &26	 \\
2454188.66520 &0.00008 &26  &2456014.4250  &0.0014  &23  &2458893.45861 	&0.00002 &26	 \\
2454188.75249 &0.00011 &26  &2456062.03196 &0.00005 &19  &2458893.54076 	&0.00002 &26	 \\
2454196.4983  &0.0009  &15  &2456063.06986 &0.00015 &19  &2458893.63166 	&0.00005 &26	 \\
2454197.360   &0.002   &15  &2456064.01443 &0.00008 &19  &2458893.71907 	&0.00002 &26	 \\
2454197.4409  &0.0006  &15  &2456064.09555 &0.00008 &19  &2458893.79966 	&0.00003 &26	 \\
2454198.3857  &0.0009  &15  &2456065.04265 &0.00009 &19  &2458893.88622 	&0.00002 &26	 \\
2454198.4748  &0.0007  &15  &2456068.05875 &0.00021 &19  &2458893.97844 	&0.00003 &26	 \\
2454202.4295  &0.0007  &15  &2456319.65508 &0.00008 &26  &2458894.05964 	&0.00002 &26	 \\
2454220.58298 &0.00009 &26  &2456319.74464 &0.00005 &26  &2458894.14297 	&0.00002 &26	 \\
2454220.66380 &0.00007 &26  &2456325.41563 &0.00019 &26  &2458894.23523 	&0.00005 &26	 \\
2454414.8946  &0.0008  &14  &2456326.45390 &0.00033 &26  &2458894.32016 	&0.00003 &26	 \\
2454414.9811  &0.0007  &14  &2456342.3651  &0.0003	&21  &2458894.40103 	&0.00003 &26	 \\
2454417.7278  &0.0006  &14  &2456342.4468  &0.0003	&21  &2458894.48951 	&0.00005 &26	 \\
2454417.9065  &0.0006  &14  &2456342.5322  &0.0009	&21  &2458894.58014 	&0.00003 &26	 \\
2454417.9863  &0.0005  &14  &2456342.6275  &0.0005	&21  &2458894.66095 	&0.00001 &26	 \\
2454422.89521 &0.00011 &26  &2456610.73911 &0.00006 &26  &2458894.74553 	&0.00002 &26	 \\
2454440.6965  &0.0010  &14  &2456610.82011 &0.00005 &26  &2458894.83840 	&0.00005 &26	 \\
2454442.7626  &0.0009  &14  &2456610.90597 &0.00009 &26  &2458894.92140 	&0.00002 &26	 \\
2454442.8523  &0.0011  &14  &2456615.72836 &0.00009 &26  &2458895.00300 	&0.00002 &26	 \\
2454442.9349  &0.0013  &14  &2456615.80960 &0.00007 &26  &2458895.09339 	&0.00006 &26	 \\
2454451.6250  &0.0012  &14  &2456619.77091 &0.00006 &26  &2458895.18162 	&0.00002 &26	 \\
2454460.6520  &0.0008  &14  &2456620.71765 &0.00003 &26  &2458895.26206 	&0.00002 &26	 \\
2454460.7440  &0.0007  &14  &2456620.79864 &0.00001 &26  &2458895.34793 	&0.00003 &26	 \\
2454460.8254  &0.0005  &14  &2456620.88433 &0.00017 &26  &2458895.44038 	&0.00004 &26	 \\
2454460.9096  &0.0006  &14  &2456623.4632  &0.0035  &24  &2458895.52239 	&0.00002 &26	 \\
2454467.7962  &0.0006  &14  &2456633.70225 &0.00003 &26  &2458895.60523 	&0.00002 &26	 \\
2454468.7399  &0.0006  &14  &2456633.78654 &0.00002 &26  &2458895.69693 	&0.00006 &26	 \\
2454468.8227  &0.0007  &14  &2456642.4783  &0.0009  &24  &2458895.78285 	&0.00002 &26	 \\
2454468.9127  &0.0009  &14  &2456642.5592  &0.0006  &24  &2458895.86363 	&0.00002 &26	 \\
2454469.6822  &0.0006  &14  &2456642.6494  &0.0011  &24  &2458895.95121 	&0.00003 &26	 \\
2454469.7708  &0.0008  &14  &2456643.5068  &0.0035  &24  &2458896.04261 	&0.00004 &26	 \\
2454469.8610  &0.0006  &14  &2456655.72285 &0.00004 &26  &2458896.12321 	&0.00002 &26	 \\
2454469.9421  &0.0006  &14  &2456660.63037 &0.00004 &26  &2458896.20721 	&0.00003 &26	 \\
2454474.76018 &0.00015 &26  &2456660.71202 &0.00002 &26  &2458896.30018 	&0.00004 &26	 \\
2454474.85028 &0.00011 &26  &2456660.79912 &0.00007 &26  &2458896.38403 	&0.00001 &26	 \\
2454474.93116 &0.00009 &26  &2456660.88863 &0.00007 &26  &2458896.46539 	&0.00003 &26	 \\
2454479.83930 &0.00016 &26  &2456660.97184 &0.00002 &26  &2458896.55462 	&0.00006 &26	 \\
2454479.92088 &0.00010 &26  &2456722.47037 &0.00017 &26  &2458896.64411 	&0.00004 &26	 \\
2454506.41171 &0.00010 &26  &2456730.47608 &0.00026 &26  &2458896.72477 	&0.00003 &26	 \\
2454506.50304 &0.00012 &26  &2456732.36386 &0.00006 &26  &2458896.81005 	&0.00004 &26	 \\
2454506.5894  &0.0006  &16  &2456744.40612 &0.00008 &26  &2458896.90268 	&0.00006 &26	 \\
2454512.5206  &0.0002  &16  &2456998.93336 &0.00014 &26  &2458896.98505 	&0.00002 &26	 \\
2454513.38185 &0.00001 &26  &2456999.01415 &0.00026 &26  &2458897.06724 	&0.00001 &26	 \\
2454513.46540 &0.00001 &26  &2457158.40951 &0.00003 &26  &2458897.15865 	&0.00004 &26	 \\
2454513.4657  &0.0009  &16  &2457390.73645 &0.00003 &26  &2458897.24531 	&0.00003 &26	 \\
2454513.5566  &0.0009  &16  &2457392.63658 &0.00005 &26  &2458897.32620 	&0.00002 &26	 \\
2454513.55706 &0.00003 &26  &2457392.71836 &0.00001 &26  &2458897.41347 	&0.00005 &26	 \\
2454516.82745 &0.00010 &26  &2457392.80048 &0.00002 &26  &2458897.50480 	&0.00004 &26	 \\
2454524.4822  &0.0008  &16  &2457392.89186 &0.00003 &26  &2458897.50484 	&0.00002 &26	 \\
2454769.8866  &0.0010  &14  &2457392.97894 &0.00002 &26  &2458897.58612 	&0.00002 &26	 \\
2454770.8343  &0.0005  &14  &2457406.65009 &0.00005 &26  &2458897.58621 	&0.00003 &26	 \\
2454770.9159  &0.0005  &14  &2457406.74123 &0.00004 &26  &2458897.66929 	&0.00005 &26	 \\
2454781.8391  &0.0007  &14  &2457406.82499 &0.00002 &26  &2458897.66944 	&0.00002 &26	 \\
2454781.9308  &0.0010  &14  &2457406.90866 &0.00003 &26  &2458897.76197 	&0.00002 &26	 \\
2454788.8087  &0.0006  &14  &2457425.4941  &0.0010  &25  &2458897.76199 	&0.00004 &26	 \\
2454788.8919  &0.0006  &14  &2457425.5774  &0.0010  &25  &2458912.63614 	&0.00002 &26	 \\
2454788.9840  &0.0012  &14  &2457425.6596  &0.0010  &25  &2458951.34527 	&0.00005 &26	 \\
2454791.7340  &0.0006  &14  &2457437.44382 &0.00001 &26  &2458962.36194 	&0.00015 &26	 \\
2454791.8164  &0.0005  &14  &2457466.3535  &0.0009  &25  &2458962.44653 	&0.00006 &26	 \\
2454791.9087  &0.0011  &14  &              &        &    &				    &        &       \\
\enddata
\tablecomments{
  Source: (1) \citet{Pocs2001}; (2) \citet{Broglia1975}; (3)
  \citet{Rodriguez1992}; (4) \citet{Hintz1997}; (5) \citet{Agerer1999}; (6)
  \citet{Pejcha2001}; (7) \citet{Zhou2001}; (8)
  \citet{Agerer2003}; (9) \citet{Hubscher2005a}; (10)
  \citet{Hubscher2005b}; (11) \citet{Klingenberg2006}; (12) \citet{Hubscher2006};
  (13) \citet{Hubscher2007a}; (14) \citet{Samolyk2010}; (15) \citet{Hubscher2007b};
  (16)\citet{Hubscher2009}; (17) \citet{Hubscher2010}; (18)
  \citet{Hubscher2011}; (19) \citet{Niu2017}; (20) \citet{Hubscher2012};
  (21) \citet{Hubscher2013a}; (22) \citet{Hubscher2013b}; (23)
  \citet{Hubscher2013c}; (24) \citet{Hubscher2014}; (25) \citet{Hubscher2017}; (26) this work.}
\end{deluxetable*}


\begin{thebibliography}{}
\expandafter\ifx\csname natexlab\endcsname\relax\def\natexlab#1{#1}\fi
\providecommand{\url}[1]{\href{#1}{#1}}
\providecommand{\dodoi}[1]{doi:~\href{http://doi.org/#1}{\nolinkurl{#1}}}
\providecommand{\doeprint}[1]{\href{http://ascl.net/#1}{\nolinkurl{http://ascl.net/#1}}}
\providecommand{\doarXiv}[1]{\href{https://arxiv.org/abs/#1}{\nolinkurl{https://arxiv.org/abs/#1}}}

\bibitem[{Aerts(2021)}]{Aerts2021}
Aerts, C. 2021, Rev. Mod. Phys., 93, 015001,
  \dodoi{10.1103/RevModPhys.93.015001}

\bibitem[{{Agerer} {et~al.}(1999){Agerer}, {Dahm}, \& {Hubscher}}]{Agerer1999}
{Agerer}, F., {Dahm}, M., \& {Hubscher}, J. 1999, Information Bulletin on
  Variable Stars, 4712, 1

\bibitem[{{Agerer} \& {Hubscher}(2003)}]{Agerer2003}
{Agerer}, F., \& {Hubscher}, J. 2003, Information Bulletin on Variable Stars,
  5485, 1

\bibitem[{{Bailer-Jones} {et~al.}(2021){Bailer-Jones}, {Rybizki}, {Fouesneau},
  {Demleitner}, \& {Andrae}}]{Bailer-Jones2021}
{Bailer-Jones}, C.~A.~L., {Rybizki}, J., {Fouesneau}, M., {Demleitner}, M., \&
  {Andrae}, R. 2021, \aj, 161, 147, \dodoi{10.3847/1538-3881/abd806}

\bibitem[{{Bowman} {et~al.}(2021){Bowman}, {Hermans},
  {Daszy{\'n}ska-Daszkiewicz}, {Holdsworth}, {Tkachenko}, {Murphy}, {Smalley},
  \& {Kurtz}}]{Bowman2021}
{Bowman}, D.~M., {Hermans}, J., {Daszy{\'n}ska-Daszkiewicz}, J., {et~al.} 2021,
  \mnras, 504, 4039, \dodoi{10.1093/mnras/stab1124}

\bibitem[{{Bowman} {et~al.}(2016){Bowman}, {Kurtz}, {Breger}, {Murphy}, \&
  {Holdsworth}}]{Bowman2016}
{Bowman}, D.~M., {Kurtz}, D.~W., {Breger}, M., {Murphy}, S.~J., \&
  {Holdsworth}, D.~L. 2016, \mnras, 460, 1970, \dodoi{10.1093/mnras/stw1153}

\bibitem[{{Breger}(2000)}]{Breger2000}
{Breger}, M. 2000, in Astronomical Society of the Pacific Conference Series,
  Vol. 210, Delta Scuti and Related Stars, ed. M.~{Breger} \& M.~{Montgomery},
  3

\bibitem[{{Breger} \& {Montgomery}(2014)}]{Breger2014}
{Breger}, M., \& {Montgomery}, M.~H. 2014, \apj, 783, 89,
  \dodoi{10.1088/0004-637X/783/2/89}

\bibitem[{{Breger} \& {Pamyatnykh}(1998)}]{Breger1998}
{Breger}, M., \& {Pamyatnykh}, A.~A. 1998, \aap, 332, 958.
\newblock \doarXiv{astro-ph/9802076}

\bibitem[{{Breger} {et~al.}(1993){Breger}, {Stich}, {Garrido}, {Martin},
  {Jiang}, {Li}, {Hube}, {Ostermann}, {Paparo}, \& {Scheck}}]{Breger1993}
{Breger}, M., {Stich}, J., {Garrido}, R., {et~al.} 1993, \aap, 271, 482

\bibitem[{{Broglia} \& {Conconi}(1975)}]{Broglia1975}
{Broglia}, P., \& {Conconi}, P. 1975, \aaps, 22, 243

\bibitem[{{Daszy{\'n}ska-Daszkiewicz}
  {et~al.}(2022){Daszy{\'n}ska-Daszkiewicz}, {Walczak}, {Pamyatnykh}, \&
  {Szewczuk}}]{Daszynska2022}
{Daszy{\'n}ska-Daszkiewicz}, J., {Walczak}, P., {Pamyatnykh}, A.~A., \&
  {Szewczuk}, W. 2022, \mnras, 512, 3551, \dodoi{10.1093/mnras/stac646}

\bibitem[{{Eastman} {et~al.}(2010){Eastman}, {Siverd}, \&
  {Gaudi}}]{Eastman2010}
{Eastman}, J., {Siverd}, R., \& {Gaudi}, B.~S. 2010, \pasp, 122, 935,
  \dodoi{10.1086/655938}

\bibitem[{{Foreman-Mackey} {et~al.}(2013){Foreman-Mackey}, {Hogg}, {Lang}, \&
  {Goodman}}]{emcee}
{Foreman-Mackey}, D., {Hogg}, D.~W., {Lang}, D., \& {Goodman}, J. 2013, \pasp,
  125, 306, \dodoi{10.1086/670067}

\bibitem[{{Goldstein} \& {Townsend}(2020)}]{Goldstein2020}
{Goldstein}, J., \& {Townsend}, R.~H.~D. 2020, \apj, 899, 116,
  \dodoi{10.3847/1538-4357/aba748}

\bibitem[{{Handler}(2009)}]{Handler2009}
{Handler}, G. 2009, in American Institute of Physics Conference Series, Vol.
  1170, Stellar Pulsation: Challenges for Theory and Observation, ed. J.~A.
  {Guzik} \& P.~A. {Bradley}, 403--409, \dodoi{10.1063/1.3246528}

\bibitem[{{Henden} {et~al.}(2016){Henden}, {Templeton}, {Terrell}, {Smith},
  {Levine}, \& {Welch}}]{Henden2016}
{Henden}, A.~A., {Templeton}, M., {Terrell}, D., {et~al.} 2016, VizieR Online
  Data Catalog, II/336

\bibitem[{{Herwig}(2000)}]{Herwig2000}
{Herwig}, F. 2000, \aap, 360, 952.
\newblock \doarXiv{astro-ph/0007139}

\bibitem[{{Hintz} {et~al.}(1997){Hintz}, {Hintz}, \& {Joner}}]{Hintz1997}
{Hintz}, E., {Hintz}, M.~L., \& {Joner}, M.~D. 1997, \pasp, 109, 1073,
  \dodoi{10.1086/133977}

\bibitem[{{Holdsworth} {et~al.}(2014){Holdsworth}, {Smalley}, {Gillon},
  {Clubb}, {Southworth}, {Maxted}, {Anderson}, {Barros}, {Collier Cameron},
  {Delrez}, {Faedi}, {Haswell}, {Hellier}, {Horne}, {Jehin}, {Norton},
  {Pollacco}, {Skillen}, {Smith}, {West}, \& {Wheatley}}]{Holdsworth2014}
{Holdsworth}, D.~L., {Smalley}, B., {Gillon}, M., {et~al.} 2014, \mnras, 439,
  2078, \dodoi{10.1093/mnras/stu094}

\bibitem[{{Hubscher}(2005)}]{Hubscher2005a}
{Hubscher}, J. 2005, Information Bulletin on Variable Stars, 5643, 1

\bibitem[{{Hubscher}(2007)}]{Hubscher2007b}
---. 2007, Information Bulletin on Variable Stars, 5802, 1

\bibitem[{{Hubscher}(2013)}]{Hubscher2013a}
---. 2013, Information Bulletin on Variable Stars, 6084, 1

\bibitem[{{Hubscher}(2014)}]{Hubscher2014}
---. 2014, Information Bulletin on Variable Stars, 6118, 1

\bibitem[{{Hubscher}(2017)}]{Hubscher2017}
---. 2017, Information Bulletin on Variable Stars, 6196, 1,
  \dodoi{10.22444/IBVS.6196}

\bibitem[{{Hubscher} {et~al.}(2013){Hubscher}, {Braune}, \&
  {Lehmann}}]{Hubscher2013c}
{Hubscher}, J., {Braune}, W., \& {Lehmann}, P.~B. 2013, Information Bulletin on
  Variable Stars, 6048, 1

\bibitem[{{Hubscher} \& {Lehmann}(2012)}]{Hubscher2012}
{Hubscher}, J., \& {Lehmann}, P.~B. 2012, Information Bulletin on Variable
  Stars, 6026, 1

\bibitem[{{Hubscher} \& {Lehmann}(2013)}]{Hubscher2013b}
---. 2013, Information Bulletin on Variable Stars, 6070, 1

\bibitem[{{Hubscher} {et~al.}(2010){Hubscher}, {Lehmann}, {Monninger},
  {Steinbach}, \& {Walter}}]{Hubscher2010}
{Hubscher}, J., {Lehmann}, P.~B., {Monninger}, G., {Steinbach}, H.-M., \&
  {Walter}, F. 2010, Information Bulletin on Variable Stars, 5918, 1

\bibitem[{{Hubscher} \& {Monninger}(2011)}]{Hubscher2011}
{Hubscher}, J., \& {Monninger}, G. 2011, Information Bulletin on Variable
  Stars, 5959, 1

\bibitem[{{Hubscher} {et~al.}(2005){Hubscher}, {Paschke}, \&
  {Walter}}]{Hubscher2005b}
{Hubscher}, J., {Paschke}, A., \& {Walter}, F. 2005, Information Bulletin on
  Variable Stars, 5657, 1

\bibitem[{{Hubscher} {et~al.}(2006){Hubscher}, {Paschke}, \&
  {Walter}}]{Hubscher2006}
---. 2006, Information Bulletin on Variable Stars, 5731, 1

\bibitem[{{Hubscher} {et~al.}(2009){Hubscher}, {Steinbach}, \&
  {Walter}}]{Hubscher2009}
{Hubscher}, J., {Steinbach}, H.-M., \& {Walter}, F. 2009, Information Bulletin
  on Variable Stars, 5874, 1

\bibitem[{{Hubscher} \& {Walter}(2007)}]{Hubscher2007a}
{Hubscher}, J., \& {Walter}, F. 2007, Information Bulletin on Variable Stars,
  5761, 1

\bibitem[{{Jenkins} {et~al.}(2016){Jenkins}, {Twicken}, {McCauliff},
  {Campbell}, {Sanderfer}, {Lung}, {Mansouri-Samani}, {Girouard}, {Tenenbaum},
  {Klaus}, {Smith}, {Caldwell}, {Chacon}, {Henze}, {Heiges}, {Latham},
  {Morgan}, {Swade}, {Rinehart}, \& {Vanderspek}}]{Jenkins2016}
{Jenkins}, J.~M., {Twicken}, J.~D., {McCauliff}, S., {et~al.} 2016, in Society
  of Photo-Optical Instrumentation Engineers (SPIE) Conference Series, Vol.
  9913, Software and Cyberinfrastructure for Astronomy IV, ed. G.~{Chiozzi} \&
  J.~C. {Guzman}, 99133E, \dodoi{10.1117/12.2233418}

\bibitem[{{Kepler} {et~al.}(2021){Kepler}, {Winget}, {Vanderbosch},
  {Castanheira}, {Hermes}, {Bell}, {Mullally}, {Romero}, {Montgomery},
  {DeGennaro}, {Winget}, {Chandler}, {Jeffery}, {Fritzen}, {Williams}, {Chote},
  \& {Zola}}]{Kepler2021}
{Kepler}, S.~O., {Winget}, D.~E., {Vanderbosch}, Z.~P., {et~al.} 2021, \apj,
  906, 7, \dodoi{10.3847/1538-4357/abc626}

\bibitem[{{Klingenberg} {et~al.}(2006){Klingenberg}, {Dvorak}, \&
  {Robertson}}]{Klingenberg2006}
{Klingenberg}, G., {Dvorak}, S.~W., \& {Robertson}, C.~W. 2006, Information
  Bulletin on Variable Stars, 5701, 1

\bibitem[{{Magic} {et~al.}(2010){Magic}, {Serenelli}, {Weiss}, \&
  {Chaboyer}}]{Magic2010}
{Magic}, Z., {Serenelli}, A., {Weiss}, A., \& {Chaboyer}, B. 2010, \apj, 718,
  1378, \dodoi{10.1088/0004-637X/718/2/1378}

\bibitem[{{Montgomery} \& {Odonoghue}(1999)}]{Montgomery1999}
{Montgomery}, M.~H., \& {Odonoghue}, D. 1999, Delta Scuti Star Newsletter, 13,
  28

\bibitem[{{Niu} {et~al.}(2013){Niu}, {Fu}, \& {Zong}}]{Niu2013}
{Niu}, J.-S., {Fu}, J.-N., \& {Zong}, W.-K. 2013, \raa, 13, 1181,
  \dodoi{10.1088/1674-4527/13/10/004}

\bibitem[{{Niu} \& {Li}(2018)}]{Niu201801}
{Niu}, J.-S., \& {Li}, T. 2018, \prd, 97, 023015,
  \dodoi{10.1103/PhysRevD.97.023015}

\bibitem[{{Niu} {et~al.}(2018){Niu}, {Li}, {Ding}, {Zhu}, {Xue}, \&
  {Wang}}]{Niu201802}
{Niu}, J.-S., {Li}, T., {Ding}, R., {et~al.} 2018, \prd, 97, 083012,
  \dodoi{10.1103/PhysRevD.97.083012}

\bibitem[{{Niu} {et~al.}(2019){Niu}, {Li}, \& {Xue}}]{Niu2019}
{Niu}, J.-S., {Li}, T., \& {Xue}, H.-F. 2019, \apj, 873, 77,
  \dodoi{10.3847/1538-4357/ab0420}

\bibitem[{{Niu} \& {Xue}(2021)}]{Niu2021}
{Niu}, J.-S., \& {Xue}, H.-F. 2021, arXiv e-prints, arXiv:2102.10259.
\newblock \doarXiv{2102.10259}

\bibitem[{{Niu} {et~al.}(2017){Niu}, {Fu}, {Li}, {Yang}, {Zong}, {Xue},
  {Zhang}, {Liu}, {Du}, \& {Zuo}}]{Niu2017}
{Niu}, J.-S., {Fu}, J.-N., {Li}, Y., {et~al.} 2017, \mnras, 467, 3122,
  \dodoi{10.1093/mnras/stx125}

\bibitem[{{Paxton} {et~al.}(2011){Paxton}, {Bildsten}, {Dotter}, {Herwig},
  {Lesaffre}, \& {Timmes}}]{Paxton2011}
{Paxton}, B., {Bildsten}, L., {Dotter}, A., {et~al.} 2011, \apjs, 192, 3,
  \dodoi{10.1088/0067-0049/192/1/3}

\bibitem[{{Paxton} {et~al.}(2013){Paxton}, {Cantiello}, {Arras}, {Bildsten},
  {Brown}, {Dotter}, {Mankovich}, {Montgomery}, {Stello}, {Timmes}, \&
  {Townsend}}]{Paxton2013}
{Paxton}, B., {Cantiello}, M., {Arras}, P., {et~al.} 2013, \apjs, 208, 4,
  \dodoi{10.1088/0067-0049/208/1/4}

\bibitem[{{Paxton} {et~al.}(2015){Paxton}, {Marchant}, {Schwab}, {Bauer},
  {Bildsten}, {Cantiello}, {Dessart}, {Farmer}, {Hu}, {Langer}, {Townsend},
  {Townsley}, \& {Timmes}}]{Paxton2015}
{Paxton}, B., {Marchant}, P., {Schwab}, J., {et~al.} 2015, \apjs, 220, 15,
  \dodoi{10.1088/0067-0049/220/1/15}

\bibitem[{{Paxton} {et~al.}(2018){Paxton}, {Schwab}, {Bauer}, {Bildsten},
  {Blinnikov}, {Duffell}, {Farmer}, {Goldberg}, {Marchant}, {Sorokina},
  {Thoul}, {Townsend}, \& {Timmes}}]{Paxton2018}
{Paxton}, B., {Schwab}, J., {Bauer}, E.~B., {et~al.} 2018, \apjs, 234, 34,
  \dodoi{10.3847/1538-4365/aaa5a8}

\bibitem[{{Paxton} {et~al.}(2019){Paxton}, {Smolec}, {Schwab}, {Gautschy},
  {Bildsten}, {Cantiello}, {Dotter}, {Farmer}, {Goldberg}, {Jermyn}, {Kanbur},
  {Marchant}, {Thoul}, {Townsend}, {Wolf}, {Zhang}, \& {Timmes}}]{Paxton2019}
{Paxton}, B., {Smolec}, R., {Schwab}, J., {et~al.} 2019, \apjs, 243, 10,
  \dodoi{10.3847/1538-4365/ab2241}

\bibitem[{{Pejcha} {et~al.}(2001){Pejcha}, {Havlik}, \& {Kral}}]{Pejcha2001}
{Pejcha}, O., {Havlik}, T., \& {Kral}, L. 2001, Information Bulletin on
  Variable Stars, 5080, 1

\bibitem[{{Petersen} \& {Christensen-Dalsgaard}(1996)}]{Petersen1996}
{Petersen}, J.~O., \& {Christensen-Dalsgaard}, J. 1996, \aap, 312, 463

\bibitem[{{Petersen} \& {Christensen-Dalsgaard}(1999)}]{Petersen1999}
---. 1999, \aap, 352, 547

\bibitem[{{Pigulski} {et~al.}(2022){Pigulski}, {Kotysz}, \&
  {Ko{\l}aczek-Szyma{\'n}ski}}]{Pigulski2022}
{Pigulski}, A., {Kotysz}, K., \& {Ko{\l}aczek-Szyma{\'n}ski}, P.~A. 2022, \aap,
  663, A62, \dodoi{10.1051/0004-6361/202243293}

\bibitem[{{P{\'o}cs} \& {Szeidl}(2001)}]{Pocs2001}
{P{\'o}cs}, M.~D., \& {Szeidl}, B. 2001, \aap, 368, 880,
  \dodoi{10.1051/0004-6361:20010043}

\bibitem[{{Poretti} {et~al.}(2005){Poretti}, {Su{\'a}rez}, {Niarchos},
  {Gazeas}, {Manimanis}, {van Cauteren}, {Lampens}, {Wils}, {Alonso}, {Amado},
  {Belmonte}, {Butterworth}, {Martignoni}, {Mart{\'\i}n-Ruiz}, {Moskalik}, \&
  {Robertson}}]{Poretti2005}
{Poretti}, E., {Su{\'a}rez}, J.~C., {Niarchos}, P.~G., {et~al.} 2005, \aap,
  440, 1097, \dodoi{10.1051/0004-6361:20053463}

\bibitem[{{Ricker} {et~al.}(2015){Ricker}, {Winn}, {Vanderspek}, {Latham},
  {Bakos}, {Bean}, {Berta-Thompson}, {Brown}, {Buchhave}, {Butler}, {Butler},
  {Chaplin}, {Charbonneau}, {Christensen-Dalsgaard}, {Clampin}, {Deming},
  {Doty}, {De Lee}, {Dressing}, {Dunham}, {Endl}, {Fressin}, {Ge}, {Henning},
  {Holman}, {Howard}, {Ida}, {Jenkins}, {Jernigan}, {Johnson}, {Kaltenegger},
  {Kawai}, {Kjeldsen}, {Laughlin}, {Levine}, {Lin}, {Lissauer}, {MacQueen},
  {Marcy}, {McCullough}, {Morton}, {Narita}, {Paegert}, {Palle}, {Pepe},
  {Pepper}, {Quirrenbach}, {Rinehart}, {Sasselov}, {Sato}, {Seager},
  {Sozzetti}, {Stassun}, {Sullivan}, {Szentgyorgyi}, {Torres}, {Udry}, \&
  {Villasenor}}]{Ricker2015}
{Ricker}, G.~R., {Winn}, J.~N., {Vanderspek}, R., {et~al.} 2015, Journal of
  Astronomical Telescopes, Instruments, and Systems, 1, 014003,
  \dodoi{10.1117/1.JATIS.1.1.014003}

\bibitem[{{Rodriguez} {et~al.}(1992){Rodriguez}, {Rolland}, {Lopez de Coca},
  {Garcia-Lobo}, \& {Sedano}}]{Rodriguez1992}
{Rodriguez}, E., {Rolland}, A., {Lopez de Coca}, P., {Garcia-Lobo}, E., \&
  {Sedano}, J.~L. 1992, \aaps, 93, 189

\bibitem[{{Samolyk}(2010)}]{Samolyk2010}
{Samolyk}, G. 2010, \jaavso, 38, 12

\bibitem[{{Schlafly} \& {Finkbeiner}(2011)}]{Schlafly2011}
{Schlafly}, E.~F., \& {Finkbeiner}, D.~P. 2011, \apj, 737, 103,
  \dodoi{10.1088/0004-637X/737/2/103}

\bibitem[{{Stassun} {et~al.}(2019){Stassun}, {Oelkers}, {Paegert}, {Torres},
  {Pepper}, {De Lee}, {Collins}, {Latham}, {Muirhead}, {Chittidi},
  {Rojas-Ayala}, {Fleming}, {Rose}, {Tenenbaum}, {Ting}, {Kane}, {Barclay},
  {Bean}, {Brassuer}, {Charbonneau}, {Ge}, {Lissauer}, {Mann}, {McLean},
  {Mullally}, {Narita}, {Plavchan}, {Ricker}, {Sasselov}, {Seager}, {Sharma},
  {Shiao}, {Sozzetti}, {Stello}, {Vanderspek}, {Wallace}, \&
  {Winn}}]{Stassun2019}
{Stassun}, K.~G., {Oelkers}, R.~J., {Paegert}, M., {et~al.} 2019, \aj, 158,
  138, \dodoi{10.3847/1538-3881/ab3467}

\bibitem[{{Torres}(2010)}]{Torres2010}
{Torres}, G. 2010, \aj, 140, 1158, \dodoi{10.1088/0004-6256/140/5/1158}

\bibitem[{{Townsend} {et~al.}(2018){Townsend}, {Goldstein}, \&
  {Zweibel}}]{Townsend2018}
{Townsend}, R.~H.~D., {Goldstein}, J., \& {Zweibel}, E.~G. 2018, \mnras, 475,
  879, \dodoi{10.1093/mnras/stx3142}

\bibitem[{{Townsend} \& {Teitler}(2013)}]{Townsend2013}
{Townsend}, R.~H.~D., \& {Teitler}, S.~A. 2013, \mnras, 435, 3406,
  \dodoi{10.1093/mnras/stt1533}

\bibitem[{{Uytterhoeven} {et~al.}(2011){Uytterhoeven}, {Moya},
  {Grigahc{\`e}ne}, {Guzik}, {Guti{\'e}rrez-Soto}, {Smalley}, {Handler},
  {Balona}, {Niemczura}, {Fox Machado}, {Benatti}, {Chapellier}, {Tkachenko},
  {Szab{\'o}}, {Su{\'a}rez}, {Ripepi}, {Pascual}, {Mathias},
  {Mart{\'\i}n-Ru{\'\i}z}, {Lehmann}, {Jackiewicz}, {Hekker}, {Gruberbauer},
  {Garc{\'\i}a}, {Dumusque}, {D{\'\i}az-Fraile}, {Bradley}, {Antoci}, {Roth},
  {Leroy}, {Murphy}, {De Cat}, {Cuypers}, {Kjeldsen}, {Christensen-Dalsgaard},
  {Breger}, {Pigulski}, {Kiss}, {Still}, {Thompson}, \& {van
  Cleve}}]{Uytterhoeven2011}
{Uytterhoeven}, K., {Moya}, A., {Grigahc{\`e}ne}, A., {et~al.} 2011, \aap, 534,
  A125, \dodoi{10.1051/0004-6361/201117368}

\bibitem[{{Wils} {et~al.}(2008){Wils}, {Rozakis}, {Kleidis}, {Hambsch}, \&
  {Bernhard}}]{Wils2008}
{Wils}, P., {Rozakis}, I., {Kleidis}, S., {Hambsch}, F.~J., \& {Bernhard}, K.
  2008, \aap, 478, 865, \dodoi{10.1051/0004-6361:20078992}

\bibitem[{{Wu} \& {Li}(2016)}]{Wu2016}
{Wu}, T., \& {Li}, Y. 2016, \apjl, 818, L13,
  \dodoi{10.3847/2041-8205/818/1/L13}

\bibitem[{{Wu} \& {Li}(2017)}]{Wu2017}
---. 2017, \apj, 846, 41, \dodoi{10.3847/1538-4357/aa8361}

\bibitem[{{Wu} \& {Li}(2018)}]{Wu2018}
---. 2018, \mnras, 478, 3871, \dodoi{10.1093/mnras/sty1347}

\bibitem[{{Wu} \& {Li}(2019)}]{Wu2019}
---. 2019, \apj, 881, 86, \dodoi{10.3847/1538-4357/ab2ad8}

\bibitem[{{Xue} \& {Niu}(2020)}]{Xue2020}
{Xue}, H.-F., \& {Niu}, J.-S. 2020, \apj, 904, 5,
  \dodoi{10.3847/1538-4357/abbc12}

\bibitem[{{Xue} {et~al.}(2018){Xue}, {Fu}, {Fox-Machado}, {Shi}, {Zhou},
  {Zhang}, {Michel}, {Yan}, {Niu}, {Zong}, {Su}, {Castro}, {Ayala-Loera}, \&
  {Altamirano-D{\'e}vora}}]{Xue2018}
{Xue}, H.-F., {Fu}, J.-N., {Fox-Machado}, L., {et~al.} 2018, \apj, 861, 96,
  \dodoi{10.3847/1538-4357/aac9c5}

\bibitem[{{Yang} {et~al.}(2012){Yang}, {Fu}, \& {Zha}}]{Yang2012}
{Yang}, X.~H., {Fu}, J.~N., \& {Zha}, Q. 2012, \aj, 144, 92,
  \dodoi{10.1088/0004-6256/144/4/92}

\bibitem[{{Zhang} {et~al.}(2020){Zhang}, {Li}, {Wu}, \& {Su}}]{Zhang2020}
{Zhang}, X., {Li}, Y., {Wu}, T., \& {Su}, J. 2020, \mnras, 494, 511,
  \dodoi{10.1093/mnras/staa667}

\bibitem[{{Zhou}(2001)}]{Zhou2001}
{Zhou}, A.~Y. 2001, \aap, 374, 235, \dodoi{10.1051/0004-6361:20010722}

\bibitem[{{Zong} {et~al.}(2016){Zong}, {Charpinet}, {Vauclair}, {Giammichele},
  \& {Van Grootel}}]{Zong2016}
{Zong}, W., {Charpinet}, S., {Vauclair}, G., {Giammichele}, N., \& {Van
  Grootel}, V. 2016, \aap, 585, A22, \dodoi{10.1051/0004-6361/201526300}

\end{thebibliography}


\label{lastpage}

\end{CJK*}
\end{document}